\documentclass[sigconf,nonacm,screen]{acmart}

\usepackage{booktabs}
\usepackage{multirow}
\usepackage{float} 
\usepackage{colortbl}
\usepackage{calc} 
\usepackage{paralist} 
\usepackage{threeparttable} 
\usepackage{makecell} 
\usepackage{balance}
\usepackage{subfig}
\usepackage{afterpage}
\usepackage{placeins}

\definecolor{htmllightgray}{HTML}{EEEEEE}
\newcolumntype{g}{>{\columncolor{htmllightgray}}r}

\definecolor{darkgray}{gray}{0.3}

\newcommand{\ie}{i.\,e.}
\newcommand{\eg}{e.\,g.}

\begin{document}

\title{Privacy Rarely Considered: Exploring Considerations in the Adoption of Third-Party Services by Websites}

\author{Christine Utz}
\orcid{0000-0003-4346-6911}
\affiliation{%
    \institution{CISPA Helmholtz Center for Information Security}
    \city{Saarbr\"{u}cken}
    \country{Germany}
}
\email{christine.utz@cispa.de}

\author{Sabrina Amft}
\affiliation{%
    \institution{CISPA Helmholtz Center for Information Security}
    \city{Hannover}
    \country{Germany}
}
\email{sabrina.amft@cispa.de}

\author{Martin Degeling}
\orcid{0000-0001-7048-781X}
\affiliation{%
    \institution{Ruhr University Bochum}
    \city{Bochum}
    \country{Germany}
}
\email{martin.degeling@rub.de}

\author{Thorsten Holz}
\orcid{0000-0002-2783-1264}
\affiliation{%
    \institution{CISPA Helmholtz Center for Information Security}
    \city{Saarbr\"{u}cken}
    \country{Germany}
}
\email{holz@cispa.de}

\author{Sascha Fahl}
\orcid{0000-0002-5644-3316}
\affiliation{%
    \institution{CISPA Helmholtz Center for Information Security}
    \city{Hannover}
    \country{Germany}
}
\email{fahl@cispa.de}

\author{Florian Schaub}
\orcid{0000-0003-1039-7155}
\affiliation{%
    \institution{University of Michigan \\School of Information}
    \city{Ann Arbor}
    \state{Michigan}
    \country{USA}
}
\email{fschaub@umich.edu}


\begin{abstract}
Modern websites frequently use and embed third-party services to facilitate web development, connect to social media, or for monetization. This often introduces privacy issues as the inclusion of third-party services on a website can allow the third party to collect personal data about the website's visitors. While the prevalence and mechanisms of third-party web tracking have been widely studied, little is known about the decision processes that lead to websites using third-party functionality and whether efforts are being made to protect their visitors' privacy.

We report results from an online survey with 395 participants involved in the creation and maintenance of websites. For ten common website functionalities we investigated if privacy has played a role in decisions about how the functionality is integrated, if specific efforts for privacy protection have been made during integration, and to what degree people are aware of data collection through third parties. We find that ease of integration drives third-party adoption but visitor privacy is considered if there are legal requirements or respective guidelines. Awareness of data collection and privacy risks is higher if the collection is directly associated with the purpose for which the third-party service is used. 
\end{abstract}


\maketitle

\section{Introduction}
\label{sec:introduction}
Contemporary websites often use third-party services for certain functionality, design, or media resources. 
The underlying reasons are as multifaceted as the purposes for which external resources are used in web development. Web content is often monetized via online advertising and marketing~\cite{libert_news_2018}, which frequently involves the inclusion of advertising networks to target ads to website visitors' presumed interests and web analytics to measure the success of online marketing campaigns. User expectations regarding the look and functionality of websites, paired with time and resource constraints in web development, were also found to drive the adoption of third-party resources~\cite{edri_ethicalwebdev_2020}, such as design frameworks, contact forms, and external media hosting. 
This reliance on third parties can come at the cost of website visitors' privacy. By embedding external resources, websites provide third-party vendors with the opportunity to collect personal data about the website's visitors, such as their IP address, visited pages, and access to long-term identifiers the third party may have stored in visitors' browsers~\cite{libert_news_2018}. This data collection potentially allows them to track people across the Web, learn large shares of their browsing histories, and use this information to infer interests or demographics. 

Considering that third-party resources are often automatically retrieved in the background without visible indication, this may be at odds with privacy legislation. For example, the European Union's General Data Protection Regulation (GDPR)~\cite{gdpr_2016}, in effect since 2018, demands that processing of personal data is grounded on one of six legal bases, including user consent, is transparently communicated, and a ``privacy by design and by default'' approach is followed.
Privacy risks of third-party website resources have been pointed out by courts and technical guides, noting, for example, that use of the most prevalent third-party service~\cite{karaj_whotracksme_2019, hulce_webalmanac_2019, englehardt_1million_2016}, Google Analytics, is only compliant with privacy law with IP anonymization~\cite{strack_anonymizeip_2020}. 
Recent years have also seen the introduction of more privacy-friendly ways to embed externally hosted media or social media functionality~\cite{heise_embetty_2020, heise_shariff_2019}. Still, post-GDPR measurements have shown little change in the prevalence of third-party web tracking~\cite{sorensen_gdprthirdparties_2019, urban_adnetworks_2020, degeling_gdpr_2019}, and practices that are already ``quite pervasive''~\cite{edri_ethicalwebdev_2020} may be hard to change.
In early 2022, several European courts and data protection authorities have directed attention towards the privacy implications of third-party use through decisions that declared the use of certain services a GDPR violation: the Austrian and French data protection boards for Google Analytics~\cite{noyb_austriaga_2022, pollet_cnilga_2022}, the Belgian one for IAB Europe's Transparency and Consent Framework (TCF), the basis for many third-party consent providers~\cite{bryant_belgiumiab_2022}, and a German court for Google Fonts
~\cite{lakshmanan_googlefonts_2022}, with more 
decisions expected to follow~\cite{burgess_beginning_2022}.

Website creators are a crucial part of the third-party tracking ecosystem, as it is them who integrate third parties into websites and enable them to track visitors' behavior across the Web.
Thus, the lack of change in third-party use on websites under the GDPR raises the question to what extent people tasked with the creation and maintenance of websites are \emph{aware} of the privacy risks of third-party use and if visitors' privacy is considered both in the decision that leads to the \emph{selection} of third-party services and in \emph{integration} itself. Though prior work has studied the history~\cite{lerner_internetjones_2016, wambach_evolution_2016} and prevalence~\cite{englehardt_1million_2016} of third-party web tracking and its underlying mechanisms, little is known about the decision processes behind the use of third-party services on websites and if website visitors' privacy is considered in the process.

Previous work that has studied developer behavior in adopting~\cite{polese_app3ps_2022} and updating~\cite{salza_update_2018, salza_update_2020} third-party libraries focused on smartphone apps, \eg, investigating developers' privacy considerations in their use of mobile advertising networks~\cite{mhaidli_ad_networks_2019, tahaei_appdevs_2022}, their awareness of data collection through third-party tools for unspecified types of functionality including ads and analytics~\cite{balebako_behaviors_2014}, and their adoption of alternative APIs that preserve location privacy~\cite{jain_apis_2014}. 
Third-party services and libraries for websites differ from those for the mobile ecosystem in their availability for a greater variety of purposes, the potential for higher technical complexity, and higher sophistication of advertising ecosystems~
\cite{leung_app_2016, urban_frontpage_2020, ikram_implicit_2019}. Websites also lack apps' distribution through a centralized platform, whose requirements may shape developers' understanding of privacy aspects, including what data is considered sensitive~\cite{tahaei_stackoverflow_2020}. 
On the Web, the omnipresence of consent notices that implement IAB Europe's TCF~\cite{hils_cmp_2020} and often list a site's third-party vendors could have led to higher awareness of data collection through third parties on websites compared to the mobile space, where consent prompts are much less prevalent
~\cite{kollnig_appconsent_2021}.

In this work, we address this research gap with findings from a mixed-methods online study with 395 participants involved in the design, development, deployment, maintenance, or management of websites. We combine survey answers with web privacy measurements and investigate how ten website functionalities frequently associated with use of third-party services have been integrated into websites and how visitors' privacy was considered in the process. 
We go beyond prior work by exploring privacy considerations between different types of functionality that may not be equally prone to third-party use~\cite{libert_news_2018}, as well as factors that influence the adoption of first- vs. third-party solutions to integrate a functionality.
More specifically, we make the following contributions: 

\begin{itemize}
    \item We extend web privacy research on the prevalence of third-party services by contrasting their use with first-party integrations for different purposes, regarding their prevalence, factors that drive use of first vs. third-party solutions, and consideration of alternatives. 
    We find that the decision in favor of third-party services, as in the mobile domain~\cite{salza_update_2020}, is driven by ease of integration, features, cost, and familiarity with a service, while privacy rarely is a decisive factor. 
    However, we find use of privacy-friendly integration for web analytics and programming/design resources, and self-hosting tends to be the primarily considered alternative to third-party solutions, rather than another third party. 
    \item Like work on cryptographic APIs~\cite{acar_apis_2017} and mobile ad networks~\cite{mhaidli_ad_networks_2019}, we find that changes to a service's default configuration are rarely reported. However, participants who did adjust defaults often did so in response to privacy-related court rulings or guidelines by data protection authorities. 
    \item We find higher awareness of data collection pertaining to a third-party service's core functionality, such as financial information for payment or behavioral data for analytics, whereas awareness is lacking for data collected in less prominent contexts, particularly the transmission of IP addresses and device information. 
    \item From a methodological perspective we contribute to the ongoing discussion about ethics in security and privacy research by discussing implications and lessons learned from using public GitHub data to recruit people involved with web development, a method previously used by developer-centered research~\cite{acar_apis_2017, acar_github_2017, nadi_apis_2016, gorski_warnings_2018, salza_update_2020, senarath_minimization_2018, senarath_embedprivacy_2018, tahaei_nudging_2021, tahaei_staticanalysis_2021, wijayarathna_passwords_2018}.
\end{itemize}

Our findings show the need for researchers and the web development community to raise awareness of the privacy risks associated with third-party use on websites, as well as the need for clearer regulatory guidance and requirements for privacy-friendly defaults.

\section{Third-Party Services in Web Development}
\label{sec:background}
Advantages of third-party use in web development differ by actor:
Web developers benefit from ease of integration as often all that is required is to copy and paste HTML or JavaScript snippets from the vendor's website~\cite{roberts_staticassets_2019}; potentially faster website load times through use of content delivery networks (CDNs) or caching in visitors' browsers if widely used~\cite{pollard_googlefonts_2020, roberts_staticassets_2019}; and the fact that many popular third-party services are available free of charge. 
The latter often comes at the cost of the third-party vendor collecting data about the website's visitors for monetization through advertising~\cite{libert_news_2018, edri_ethicalwebdev_2020}. 
Independent of the functionality a third-party service provides to the website, requesting a remotely hosted resource via HTTP inherently involves the transmission of the website visitors' IP address, which some jurisdictions consider personal information~\cite{ec_personaldata_2022}, to the remote server, along with device information in the browser's user agent and the currently viewed page. The third party can use these to infer additional information about individuals, such as other websites they visit that also include the third-party service~\cite{libert_hidden_2015}. A mitigation is to host the remote resource locally, if possible~\cite{edri_ethicalwebdev_2020, libert_news_2018}.

Other privacy risks and mitigations depend on the type of functionality provided. As our study is centered around common use cases for third-party services in web development, we started by identifying these through review and comparison of existing categorizations in the literature and by web tracking projects. 
We found such classifications in the works of S{\o}rensen and Kosta~\cite{sorensen_gdprthirdparties_2019}, Libert and Nielsen~\cite{libert_news_2018}, by WhoTracks.me~\cite{karaj_whotracksme_2018}, Third Party Web~\cite{hulce_thirdpartyweb_2021, hulce_webalmanac_2019}, and DuckDuckGo's Tracker Radar~\cite{duckduckgo_trackerradar_2021}. 
While categorizations differ in granularity and focus, we identified large overlap from the perspective of website owners. We did not consider categories that apply only in a first-party context (\eg, hosting, distribution) or only make sense combined with other categories (\eg, tag management). 
We ended up with ten common website functionalities, shown in Table~\ref{tab:3p-categories} with associated privacy risks and possible mitigations. The latter are generally possible on two levels: \emph{selection} how to integrate the desired functionality (self-implemented, locally or remotely hosted third-party service) and efforts in \emph{integration} of the selected solution to configure it in a more privacy-friendly way.

\begin{table*}[th!]
    \centering
    \footnotesize
    \caption{Categories of website functionalities included in this study for which use of third-party (3P) services is common.
    }
    \label{tab:3p-categories}
    \begin{threeparttable}
    \setlength\tabcolsep{4pt} 
    \begin{tabular}{@{}p{0.08\linewidth}p{0.25\linewidth}p{0.2\linewidth}p{0.25\linewidth}p{0.17\linewidth}@{}}
        \toprule
         \textbf{Functionality}& \textbf{Definition} & \textbf{Popular 3P solution(s)} & \textbf{Specific privacy risks\tnote{1}} & \textbf{Possible alternatives\tnote{2}} \\
         \midrule
         Advertising & Advertising for third-party goods or services to generate revenue for the website. & Google AdSense, Amazon Advertising, Criteo, Taboola, Outbrain & Targeting and profiling based on browsing behavior and device info; data sharing w. large advertising ecosystems & Static or context-based ads~\cite{edri_ethicalwebdev_2020}, affiliate links, sponsored content \\
         \midrule
         Analytics & Measurement of visitors' behavior to evaluate website performance and marketing success. & Google Analytics, Scorecard Research, New Relic, Yandex & Extensive data collection; data sharing with others (\eg, ad networks); tracking of browsing behavior across the Web due to widespread use~\cite{karaj_whotracksme_2019, hulce_webalmanac_2019, englehardt_1million_2016} & Config. to collect less data~\cite{google_gaprivacy_2021}; services that collect less data or can be self-hosted (\eg, Matomo)~\cite{edri_ethicalwebdev_2020} \\
         \midrule
         Embedded media & Non-text content (\eg, videos, audio files, slideshows, interactive maps) embedded into web pages. & Hosting: YouTube (videos), Google Maps (maps); embedding code by hosting 3P & Data transmitted upon page load, not only upon interaction with remotely hosted embedded content & \makecell[tl]{Self-hosting, two-\\click solutions~\cite{heise_embetty_2020}, \\YouTube-nocookie~\cite{edri_ethicalwebdev_2020}} \\
         \midrule
         \makecell[tl]{Customer \\interaction} & Mechanisms that enable specific website--visitor interactions (\eg, contact forms, comments, chat). & Google Forms, Facebook Comments, Disqus & Various personal data transmitted; leakage of this data to third parties, including ad networks, even before submission~\cite{senol_leakyforms_2022, starov_contact_2016}; Disqus: data sharing with ad networks by default without notice~\cite{gundersen_disqus_2019, brown_disqus_2017} & Plugins for content management system (CMS) \\
         \midrule
         User login / authentication & Allows users to create accounts on the website and log in. & Single-sign on with credentials from popular services (\eg, Apple, Google, Twitter, Facebook) & Providers can learn on which other sites people use their credentials and when~\cite{karegar_sociallogin_2020} & CMS-provided integration, privacy-friendly identity providers \\
         \midrule
         Payment & Allows visitors to pay for services and goods offered on the website. & Varies between regions~\cite{busse_payment_2020}, \eg, PayPal, Venmo, Alipay & Sharing of sensitive personal and financial information with payment provider~\cite{preibusch_shopping_2016} and possibly other 3Ps uninvolved in transaction~\cite{paypal_thirdparties_2022} & Limited by prevalence and practicality; pure 1P: cash, gift cards; only banks: direct bank transfer 
          \\
         \midrule
         \makecell[tl]{Privacy \\notices / \\forms} & Interface elements that help fulfill transparency, consent, and opt-out requirements from privacy laws (\eg, GDPR / ePrivacy Directive in EU, CCPA in US). & Consent Management Providers (CMPs) implementing compliance frameworks by the Internet Advertising Bureau (IAB) & 
         Not always correctly implemented, so visitor data is collected without prior consent~\cite{matte_iab_2020}; frequent use of dark patterns~\cite{nouwens_darkpatterns_2020, utz_consent_2019} & Self-implementation~\cite{degeling_gdpr_2019}; ensuring proper integration with critical website features \\
         \midrule
         Programming / design & Programming frameworks and design resources (\eg, web fonts, CSS / JS libraries). & Google Fonts, jQuery, Bootstrap & \makecell[tl]{Only general risks\tnote{1}}  
         & Self-hosting~\cite{roberts_staticassets_2019, pollard_googlefonts_2020, edri_ethicalwebdev_2020} \\ 
         \midrule
         \makecell[tl]{Social \\media \\integration} & Interface elements that connect a website with social media (SM) services (\eg, link to the website's SM profile, SM share buttons, embedded SM feeds). & Code provided by SM service (\eg, Facebook, Twitter, Instagram) & Data transmission upon page load;  
         in EU, liability of site owners for data processing by SM companies through buttons/widgets~\cite{bodoni_likebutton_2019} & Limited (3P by definition required) -- two-click mechanisms~\cite{panzi_socialshareprivacy_2012, heise_shariff_2019, heise_embetty_2020}, static profile links \\
         \midrule
         \makecell[tl]{Website \\protection} & Mechanisms to protect against (distributed) denial-of-service attacks, spam, or data scraping. & \makecell[tl]{Google reCAPTCHA, services \\based on text / behavioral analy-\\sis, security proxies (Cloudflare)} & Wide range of behavioral data collected to distinguish humans from bots~\cite{oreilly_recaptcha_2015, davis_recaptcha_2019, edri_ethicalwebdev_2020} & Against non-targeted spam: honeypots, easy math or language questions~\cite{davis_recaptcha_2019} \\
         \bottomrule
    \end{tabular}
    \begin{tablenotes}
    \item[1]General risks are (i) transmission of visitors' IP address and user agent to the third-party service, which can allow the latter to track people across the Web, especially if the service is widely used~\cite{libert_hidden_2015}; and (ii) the third party potentially requiring visitors to accept extensive privacy policies~\cite{edri_ethicalwebdev_2020, davis_recaptcha_2019}. 
    \item[2]Always viable are self-implementation (except for payment and some social media integration) and using a third-party service that collects less personal information.
    \end{tablenotes}
    \end{threeparttable}
\end{table*}

\section{Related Work}
\label{sec:relatedwork}
Previous work has studied the prevalence and evolution of third-party web tracking and developers' privacy behaviors in third-party use in the mobile app ecosystem. 

\paragraph{Evolution of Third-Party Web Tracking}
\label{sec:relatedwork-webtracking}

Web tracking has been studied extensively, including the prevalence of third-party tracking services on websites. Tracking has been identified since 1996, and since then increased in prevalence and complexity~\cite{lerner_internetjones_2016}, with the most popular services covering up to 75\,\% of websites in 2015~\cite{wambach_evolution_2016} and hundreds of different known tracking services~\cite{roesner_3ptracking_2012} whose use increases with website popularity, and visible differences between regions and website types~\cite{hu_tptrackers_2020}. Large-scale investigations confirmed that more than half of websites leak user data or load third-party scripts~\cite{libert_hidden_2015}. 
The GDPR going into effect in May 2018 increased the prevalence of cookie consent notices, while actual tracking practices did not change much~\cite{degeling_gdpr_2019} or could not be directly attributed to the GDPR~\cite{sorensen_gdprthirdparties_2019}. While there were clear differences between website visits from US or European users, implying that companies collect less data from the latter~\cite{dabrowski_gdpr_2019}, previous research overall did not find significant positive changes due to the GDPR.

\paragraph{Developers' Privacy Considerations} 
\label{sec:relatedwork-devconsiderations}
Developers' considerations of users' privacy have been studied in different contexts, but there are few insights into \emph{why} specific third-party services are used in web development.
Previous work found that developers of mobile apps are often unaware of third-party data collection~\cite{balebako_behaviors_2014}, and therefore tend to collect more data than necessary. Furthermore, developers showed a limited perception of privacy threats, often based on their organization's guidelines~\cite{hadar_designers_2017}. 
Mhaidli et~al. investigated how and why mobile app developers use and choose ad networks and whether they consider  associated risks for users~\cite{mhaidli_ad_networks_2019}. They found that developers see advertisements as the only viable way to monetize their apps and consider ad networks to be responsible for protecting app users’ privacy, not themselves. Tahaei et al. confirmed this and showed that app developers find existing privacy information and controls confusing and hard to use~\cite{tahaei_appdevs_2022}.
Other studies investigated public forums to see how developers deal with privacy regulations and changes to them, finding that they mostly try to uphold standards defined by large companies~\cite{tahaei_stackoverflow_2020} or are focused on recent changes or events~\cite{li_reddit_2020} when discussing privacy. 
When asked to solve privacy-focused tasks, developers tend to use better-documented alternatives and copy examples, which could be adopted by privacy-friendly services~\cite{jain_apis_2014}. They often struggle with embedding privacy into their application due to a lack of knowledge, privacy contradicting app requirements, or task complexity~\cite{senarath_embedprivacy_2018, peixoto_devprivacy_2020}. Another problem are third-party vendors' competing business interests, leading them to employ dark patterns that steer developers towards privacy-unfriendly defaults~\cite{tahaei_devsresponsible_2021}.

\section{Method}
\label{sec:method}
To investigate the privacy practices and decision processes behind third-party use on websites, we conducted a mixed-methods study consisting of an online survey with 395 people involved in the creation and administration of websites, paired with an analysis of participants' websites, if provided in the survey.

\subsection{Survey Design}

Our survey was inspired by the work of Mhaidli et al.~\cite{mhaidli_ad_networks_2019} and consisted of five parts. It was conducted in English and implemented on a self-hosted LimeSurvey instance. 
To prevent early priming about privacy, we framed the survey as exploring practices in the selection and use of web technologies on websites and only introduced questions about privacy and data collection practices in Part~4. 
Appendix~\ref{sec:appendix-survey} contains the full survey.

Part 1 assessed participants' background regarding their work on websites, including  experience with the functionalities in Table~\ref{tab:3p-categories}.

To provide context for the rest of the survey, Part~2 asked participants to think of one specific website they had recently worked on and to only keep this website in mind for subsequent questions. Participants could optionally provide the website's URL (Q2-0). The survey consent form explained that this information would be used to check which web technologies were present on the website. At this point we investigated the methodological question if requiring participants to provide a website had an effect on dropout rates: We made Q2-0 mandatory for half of GitHub-recruited participants (see Section~\ref{sec:recruitment}) but could not find evidence that this had an impact on dropout rates or willingness to provide a website.
Part~2 proceeded to ask about website metadata, including the country it was based in, the participant's role with regard to the website, and which of the ten functionalities in Table~\ref{tab:3p-categories} were present on the site (Q2-6). To balance level of detail and survey length, we chose to display more detailed questions only for up to three functionalities. For this, Q2-7 asked, for each functionality indicated to be present in Q2-6, to what degree the participant had been involved in the decision of how this functionality should be integrated (selection), in the integration process itself, and in maintenance or management of the integrated solution. From the functionalities for which any kind of involvement had been indicated, three were randomly selected, for which Parts~3 and 4 would be shown.

Part~3 investigated how a functionality was integrated in terms of first- vs. third-party solutions and, if applicable, embedding mechanism. It also asked about the underlying decision process including reasons for selection and considered alternatives, information sources, and the people involved. 
Part 4 explored participants' understanding of the data collected through third-party services and efforts made to protect visitors' privacy in the integration process.

Finally, Part~5 asked demographic questions and if participants had received training or educated themselves on data protection or privacy. 
At the end, participants were debriefed about the study's privacy focus and given the option to either withdraw from the study or to submit their answers. Six participants withdrew here. 

To assure survey quality, we first conducted ``think-aloud'' cognitive interviews with seven web developers and two content creators, recruited via convenience sampling. 
After each interview, we addressed identified issues and repeated this process until no further issues emerged.
A pilot launch of the survey with 101 participants recruited from GitHub (see Section~\ref{sec:recruitment}) did not yield evidence of any remaining issues, so we proceeded with data collection.

\subsection{Recruitment}
\label{sec:recruitment}

Our recruitment approach was guided by the goal to obtain different perspectives on website functionality integration. We leveraged two recruitment channels to reach a diverse sample: websites' contact information to reach individuals in a range of website-related roles, and GitHub to reach web developers.
People were eligible to participate if they were at least 18 years old, worked on websites in some capacity (\eg, website design, development, deployment, maintenance, management), and were comfortable taking the survey in English. Participation was voluntary and uncompensated.

To cover a diverse range of websites in recruitment, we searched the top 100,000 popular website domains on the Tranco list\footnote{List from September 1, 2020 (\url{https://tranco-list.eu/list/64WX}).}~\cite{lepochat_tranco_2019} for email addresses related to a website's technical administration. 
We visited each domain on the Tranco 100K in October 2020 using OpenWPM 0.13~\cite{englehardt_1million_2016} and searched the homepage for links assumed to lead to subpages containing privacy policies, terms of service, and contact information. We identified these using a list of key phrases compiled through manual inspection of 10 websites randomly sampled for each of the top 20 website languages in the Tranco list. 
We downloaded the corresponding subpages and the homepage and searched them for email addresses with a regular expression. 
Since websites often list contacts responsible for the content (\eg, editors on news pages, politicians on government sites) rather than administration, we excluded subpages with more than four email addresses. 
After removing duplicates, invalid email addresses, and subpages with more than 4 addresses, we were left with 109,862 unique email addresses for 53,496 websites.

Previous work studying web developers' security and privacy practices has used public GitHub repositories to recruit developers on a large scale~\cite{acar_apis_2017, acar_github_2017, nadi_apis_2016, gorski_warnings_2018, salza_update_2020, senarath_minimization_2018, senarath_embedprivacy_2018, tahaei_nudging_2021, tahaei_staticanalysis_2021, wijayarathna_passwords_2018}. We also used this approach because it allowed us to recruit people likely involved with web development without hand-picking them, as would have been the case for one-by-one contact on platforms such as LinkedIn. 
Though prior work is not always clear on where exactly on GitHub users' email addresses were collected (options include commit email addresses and users' profile pages), from discussions with authors of some previous studies we know that the use of commit email addresses is common. 
Following this previously used method, we analyzed commits made into public GitHub repositories in August 2020 to identify e-mail addresses of people working on websites, as indicated by the respective commit including file extensions related to web development (.js, .php, .css, .html, .htm). 
Anticipating a low response rate, we sent invitations to 37,000 email addresses, in addition to 12,000 contacted during pilot testing.

\subsection{Research Ethics}
\label{sec:ethics}
Prior to conducting the study we looked into opportunities for ethical and data protection review at our institutions. At the time this study was designed, conducted, and evaluated, the authors were affiliated with Leibniz University Hannover (LUH) and Ruhr University Bochum (RUB), both located in Germany, and the University of Michigan (U-M) in the US. RUB only had an IRB for research in psychology, which was not meant to be mandatorily consulted by security and privacy researchers. LUH's IRB only targeted project proposals, not individual research papers. The co-author from U-M did not directly work with raw response data or interact with participants and confirmed with U-M's IRB that their oversight and approval was therefore not required. Nevertheless, we followed best practices for research conduct and transparency. 
To ensure GDPR compliance of our study, we consulted RUB's and LUH's data protection officers. They both independently considered our study design and specifically the approach for GitHub recruitment to be covered by the GDPR's research privilege.

In Q2-2 we required some participants to provide the URL of a website they had worked on, following Mhaidli et al.'s study design~\cite{mhaidli_ad_networks_2019}. 
We explained in the initial consent form that this data would only be used to check the website for the presence of third-party services. Participants required to fill this field were able to drop out or proceed without penalty by entering arbitrary input.

Regarding recruitment, we carefully considered the implications of sending email invitations to website contacts and GitHub developers at a large scale. 
As mentioned above, the two consulted DPOs considered this recruitment approach to be GDPR-compliant. 
We contacted each email address only once (\ie, we did not send any confirmations or reminders) and gave email recipients a one-click option to opt-out of further contact. 
Still, we received a small number of emails with negative sentiments from people who were not aware that their public GitHub commits contained their email address. 
Upon this feedback we put up a page on our institution's website that explained our study, why the GitHub-recruited recipient's email address was visible in commits into public repositories, and what steps could be taken to hide it. 
Despite these efforts, one recipient filed a complaint with our state's data protection authority, upon which we immediately stopped recruitment via GitHub, rather than waiting for the outcome. 
Three months later the DPA informed us that they did not consider the GDPR's research privilege to apply, because GitHub users, who are often unaware of their commit email addresses being publicly available, do not expect to be contacted via these addresses for the purpose of scientific research. We discuss the concrete problem with GitHub's mechanics for email addresses in more detail in Section~\ref{sec:discussion-ethics}. 
The DPA advised us to refrain from future recruitment via public GitHub commits but did not take formal action.

When we designed and launched the study, ethical concerns with recruitment via public GitHub commits were not obvious: The method was established in the community~\cite{acar_apis_2017, acar_github_2017, gorski_warnings_2018, nadi_apis_2016, senarath_embedprivacy_2018, senarath_minimization_2018,  wijayarathna_passwords_2018}, even post-GDPR~\cite{tahaei_nudging_2021, tahaei_staticanalysis_2021, salza_update_2020}, and had passed ethical or IRB review at different universities in the US, Europe, Australia, and at the NIST Human Subjects Protection Office. 
As such we followed established research practice at the time, as well as sought consultation/approval regarding GDPR from two data protection officers from different institutions, who independently concluded the recruitment method to be covered by the GDPR’s research privilege.
In hindsight, we agree with participants' and the DPA’s concerns regarding GitHub recruitment, which is why we decided to fully discuss our experience in this paper. 
We consider this aspect of our work a valuable lesson learned for the community in how legal or ethical assessment of established study methods can -- and should -- evolve. 
Section~\ref{sec:discussion-ethics} discusses implications for future work.

We want to stress that all participants whose data is reported in this paper provided their information with informed consent, obtained both at the beginning of the survey and at the end after debriefing about the study's privacy focus. The issue pointed out by the DPA lies with the recruitment method, not with the data we received from the willing and consenting survey participants.

\subsection{Data Cleaning}
\label{sec:datacleaning}
Across all recruitment phases, 2,177 people opened the survey link, 667 proceeded past the welcome page, and 452 completed the survey. 
Out of these, we removed 41 that had not seen Parts~3 and 4 due to a lack of reported involvement, nine who selected contradictory levels of involvement, and seven who provided multiple websites. 
To increase data quality, we examined response times. Average completion time was 20:42 minutes. We did not observe any suspicious patterns and thus did not remove any answers. This left us with a total of 395 valid responses.
Two authors inspected all open-response ``Other'' answers and re-coded answers that matched existing closed-ended options after discussion and mutual agreement. 
For website analysis, one author inspected all provided URLs (Q2-0) and removed all answers that were not URLs (\eg, ``client confidential'') or could not be resolved to a website.

\subsection{Data Analysis}

Two of the authors applied thematic analysis~\cite{clarke_thematic_2014} to the answers to open-ended questions. First they independently reviewed the data to identify recurring themes and created individual codebook drafts for each question. Next, they discussed these drafts and merged them into a first joint codebook. All data was then jointly coded by both researchers, who discussed problematic cases until an agreement was reached, which at times required refining codes' definitions and scopes and, thus, revisiting previously coded answers. We did not compute inter-rater reliability, as the number of responses was small enough to not require splitting up between multiple researchers~\cite{mcdonald_reliability_2019}. 
Each open-ended response could be assigned one or more codes, as participants often mentioned more than one relevant talking point. Appendix~\ref{sec:codebooks} contains the final codebooks.

To assess to which extent participants' responses about websites' integrated functionalities matched actual practice, we checked the provided websites with OpenWPM~\cite{englehardt_1million_2016}. 
For each provided URL, we accessed the front page, searched it for links to subpages, and visited up to 100 unique pages randomly selected from these to ensure we gained a complete picture~\cite{urban_frontpage_2020}. 
We performed crawls from Germany, California, and India to cover possible differences between jurisdictions~\cite{dabrowski_gdpr_2019, vaneijk_location_2019, hils_cmp_2020}. 
For each page, we collected all HTTP(S) requests and compared the list of found third-party services with those mentioned in the respective survey response, using the WhoTracks.me~\cite{karaj_whotracksme_2019} categorization as a basis. 
Finally, we compiled metadata on the provided websites: top-level domains (TLDs), website topics based on the McAfee Real-Time Database~\cite{mcafee_trustedsource}, and popularity based on the same Tranco list we used for recruitment.

For data analysis we mainly rely on descriptive statistics because the variance in response counts per website functionality would cause statistical tests to often be underpowered. Where statistical tests are appropriate and possible we used Fisher's exact tests to check if differences between categories were significant and corrected for multiple tests with the Benjamini-Hochberg procedure.

\section{Results}
\label{sec:results}
Our results show that, as in other domains, user privacy is rarely considered in web development. Yet, we do find influence of regulators' guidelines for some types of functionality, and self-hosting is a prominently considered alternative to third-party use. 
We also find a widespread lack of awareness that third-party use implies transmission of IP addresses and device metrics to the third party.

\subsection{Sample}
\label{sec:sample}

We first describe the sample of 395 participants and 361 websites they provided to support the main part of the survey.

\subsubsection{Participant Demographics and Background}

Participants predominantly identified as men (85.1\,\%; Q5-2), are most frequently in the 18--24 (33.4\,\%) or 25--34 (30.6\,\%) age ranges (Q5-1), and the majority holds a bachelor's degree (35.2\,\%; Q5-3). Most reported degrees (Q5-4) were in technical fields, with the most common non-technical degree being in business/economics (10.4\,\%).  This is consistent with demographics surveys of people working with web technologies, whose large majority are men, typically in the 24--34 age range, holding a bachelor's degree in technical fields~\cite{greif_stateofjs_2021, stackoverflow_devsurvey_2021, zippia_webdev_2022, datausa_webdev_2022}.

Participants' work with websites (Q1-2) was most frequently in a full-time position (41.8\,\%), though freelancing and part-time employment were also common, as was non-paid work (hobbyist 31.4\,\%).
In the last three years, participants had mostly worked on 2--5 websites (43.8\,\%; Q1-1). As for previous experience with the ten website functionalities (Q1-3), all but one participant reported at least one functionality, with a mean of 5.28 (sd 2.37, median 5). Experience with front-end programming or design libraries (83.0\,\%) and user login or authentication (80.5\,\%) was most common, while the fewest participants had worked with privacy plugins (29.9\,\%) and advertising (23.0\,\%).
Participants held on average 3.4 different website-specific roles (std 2.58, min 1, max 13, median 3; Q2-1) and most often worked as (web) developer, programmer, or software engineer (85.3\,\%). Other frequently reported roles include administrator/web operator, user experience design, content creator or contributor, and product or project manager. Most participants worked alone (35.7\,\%) or in teams of sizes 2--5 (35.7\,\%) (Q2-2).
42.0\,\% had received prior privacy training. The most common resources of such training were self-study (38.6\,\% of participants with training), employer training, courses at a university or school, and other non-online courses, including certifications such as CISSP.
Table~\ref{tab:participant-stats} in Appendix~\ref{sec:sample-stats} has detailed data about participants' demographics and background in their work with websites.

\subsubsection{Websites Provided by Participants}

In Q2-0, we asked participants to provide a website they had recently worked on that would serve as a reference for Parts~3 and 4 of the survey. Data cleaning left us with 361 unique valid websites, for which we compiled descriptive statistics.  
The most frequently occurring TLDs were .com, .org, and .de, followed by domains associated with web development, such as .github.io or .dev. 
Thematic classifications by McAfee were available for 264 (83.8\,\%) domains, the most common being Business, Internet Services, and Education/Reference. 
141 registered domains (44.8\,\%) appeared on the Tranco top 1-million list, with a mean ranking of 104,767 (min 5, max 958,899, std 168,620.3, median 46,695). 
Overall we find that participants mainly reported international sites aimed at providing services or information, but also a significant amount of smaller and/or personal sites hosted on popular platforms and a multitude of other thematic categories, creating a diverse sample of websites.

Participants named 72 different countries as the seat of the company behind the website (Q2-3). Coding of the open-ended answers to Q2-4 revealed that the websites were mostly targeted at a global or multi-regional audience; Table~\ref{tab:website-stats} in Appendix~\ref{sec:sample-stats} also lists the most popular individual target regions. Almost half of the websites (44.8\,\%) were reported not to have a website-specific revenue model (Q2-5). On average they relied on 0.91 sources of revenue (std 1.03, min 0, max 5, median 1). Most common were products/services sold on websites (20.5\,\%), subscriptions/membership (17.5\,\%), and revenue streams not explicitly listed in Q2-5 (14.4\,\%).

Table~\ref{tab:website-stats} in Appendix~\ref{sec:sample-stats} contains the full website statistics.

\subsection{Privacy Considerations in Selection}
\label{sec:selection}

To find out if privacy played a role in \emph{the decision how to integrate} a desired functionality, we investigated what functionalities were present on participants' websites, whether they were integrated via first- or third-party solutions, and the underlying decision process, including considered alternatives, consulted information sources, and the people involved.

\subsubsection{Integrated Functionalities}
\label{sec:integrated-functionalities}

In Q2-6 we asked participants which of the ten functionalities in Table~\ref{tab:3p-categories} were present on their website. Participants' websites used on average 5.2 of them (sd 2.3, min. 1, max. 10, median 5).
In its ``present'' column, Table~\ref{tab:categorystats} lists how often each functionality was mentioned. The numbers show that the reported prevalence of functionalities differs greatly. Most commonly used were programming or design resources (355 / 89.9\,\% of websites), customer interaction tools (268 / 67.8\,\%), and web analytics (251 / 63.5\,\%).

\begin{table}[tb]
\centering
\caption{Reported functionalities on websites (Q2-6; $ n $ = 395), participants' involvement with them (Q2-7; relative to ``present''), and, based on that, how often they were randomly assigned survey parts~3 and 4. 
}
\label{tab:categorystats}
\footnotesize
\begin{tabular}{lrggggr}
\toprule
& pre- & \multicolumn{4}{c}{\cellcolor{htmllightgray}Participants' involvement} & as- \\
{} & sent                                 &  sel. &  int. & maint. &  none & signed   \\
{}          & \textbf{n} & \textbf{\%} & \textbf{\%} & \textbf{\%} & \textbf{\%} & \textbf{n}\\
\midrule
Advertising             &       67 &       44.8 &     46.3 &     32.8 &        26.9 &     25 \\
Analytics               &      251 &       47.4 &     40.6 &     46.2 &        17.1 &    126 \\
Customer Interaction    &      268 &       53.0 &     46.6 &     45.1 &        10.8 &    138 \\
Embedded Media          &      248 &       55.6 &     48.0 &     45.2 &         9.7 &    141 \\
Login/Auth.  &      265 &       48.7 &     41.5 &     40.8 &        17.4 &    137 \\
Payment                 &      101 &       43.6 &     40.6 &     29.7 &        26.7 &     37 \\
Programming/Design    &      355 &       61.7 &     57.7 &     46.2 &         8.7 &    235 \\
Privacy   &      136 &       40.4 &     36.0 &     33.8 &        30.9 &     57 \\
Social Media            &      186 &       53.8 &     44.1 &     40.3 &        16.7 &    101 \\
Website Protection      &      187 &       51.3 &     39.0 &     39.0 &        24.6 &     70 \\
\bottomrule
\end{tabular}
\end{table}

To assess the number of third parties the websites actually use, we combined the data collected from three server locations to ensure that no configurations dependent on visitors' IP or region biased our results. Out of 361 unique websites provided we were not able to access 10. On average, each website contacted 6.2 third-party domains (min 0, max 144, std 6.95, median 3) and 80 sites made no requests to third parties at all.

For 76 sites we found mismatches between Q2-6 responses and third parties observed on the website. The most common observation was a request to Google's advertising domain \texttt{doubleclick.com} (42 cases), followed by site analytics (14), CDNs (12), customer interaction (6), and embedded media (5). The rest belonged to other functionalities not covered by the survey.
The high prevalence of requests to advertising domains despite the fact that developers had not reported the use of advertising -- confirmed by manual inspection -- can be explained by third parties loading additional services~\cite{urban_frontpage_2020}. The majority of requests went to \texttt{doubleclick.com}, contacted by locally hosted Google Analytics scripts. Other cases involved social media bookmarking services like AddThis or ShareThis that contact various advertising domains.

\begin{table}[tb]
\centering
\footnotesize
\caption{Prevalence of common third-party services used on 351 websites compared to privacy-friendly alternatives.}
\label{tab:privfriendlyhosting}
\begin{minipage}{0.48\textwidth}
\renewcommand{\thefootnote}{\thempfootnote}
\renewcommand{\arraystretch}{0.9}
\begin{tabular*}{\textwidth}{@{}>{\bfseries}ll@{\extracolsep{\fill}}*{6}{r}}
\toprule
& \textbf{Integration Solution} & \multicolumn{1}{c}{\textbf{n}} & \multicolumn{1}{c}{\textbf{\%}} \\
\midrule
\multirow{4}{*}{\rotatebox[origin=c]{90}{\textbf{Analytics}}}
& Google Analytics & 158 & 45.0 \\
& Google Analytics w/ IP anonymization & 24 & 6.8 \\
& Privacy-friendly (Matomo/Piwik) & 15 & 4.3 \\
& Only privacy-friendly & 11 & 3.1 \\
\midrule
\multirow{4}{*}{\rotatebox[origin=c]{90}{\textbf{Video}}}
& YouTube & 74 & 21.1 \\
& Vimeo & 12 & 3.4 \\
& Privacy-friendly (YouTube-nocookie) & 16 & 4.5 \\
& Only privacy-friendly & 6 & 1.7 \\
\midrule
\multirow{3}{*}{\rotatebox[origin=c]{90}{\textbf{Maps}}}
& Google Maps & 38 & 10.8 \\
& Privacy-friendly (OpenStreetMap) & 3 & 0.9 \\
& Only privacy-friendly & 2 & 0.6 \\
\midrule
\multirow{4}{*}{\rotatebox[origin=c]{90}{\textbf{Design}}}
& Google Fonts / Font Awesome & 244 & 69.5 \\
& Privacy-friendly (3P-hosted) & 6 & 1.7 \\
& Privacy-friendly (self-hosted) & 86 & 24.5  \\
& Only self-hosted fonts & 22 & 6.3 \\
\midrule
\multirow{3}{*}{\rotatebox[origin=c]{90}{\textbf{Progr.}}}
& jQuery from CDN & 72 & 20.5 \\
& Privacy-friendly (self-hosted) & 138 & 39.3 \\
& Only privacy-friendly & 101 & 28.8 \\
\bottomrule

\end{tabular*}
\end{minipage}
\end{table}

In the other direction, 136 responses reported functionalities for which website analysis did not find obvious requests to matching third parties. The majority of these cases concern scripts for customer interaction (64), embedded media (70), or social media integration (46). Besides methodological limitations outlined in Section~\ref{sec:limitations}, the explanation was often that the functionality was hosted locally, \eg, via CMS plugins, as reported in Section~\ref{sec:integrationtypes}.

Last, we compared the hosting strategies against privacy-friendly recommendations~\cite{edri_ethicalwebdev_2020}. Table~\ref{tab:privfriendlyhosting} lists results for selected services. We found that for many common third-party services like analytics, videos, and maps the main strategy was to embed the well-known services. For example, 158 websites made use of Google Analytics, while only 15 used the privacy-friendly alternative Matomo. Out of those 15 another 4 were found to be using both, \eg, on subsites. For more technical functionalities like programming and design resources we observed more variation in first- vs. third-party hosting. While we found only six websites that used privacy-friendly font hosting sites (such as Fork Awesome or Fontello~\cite{edri_ethicalwebdev_2020}), 86 hosted additional fonts on their own server. For the widely used web programming library jQuery the results were reversed: The majority (138) self-hosted the script, while 72 used CDNs to serve the files. Again there were sites using both strategies, for example, when a library was used multiple times by different components or plugins.

\subsubsection{Prevalence of First-Party vs. Third-Party Solutions}
\label{sec:integrationtypes}

\begin{figure*}[tb]
	\centering
    \includegraphics[width=1.0\textwidth]{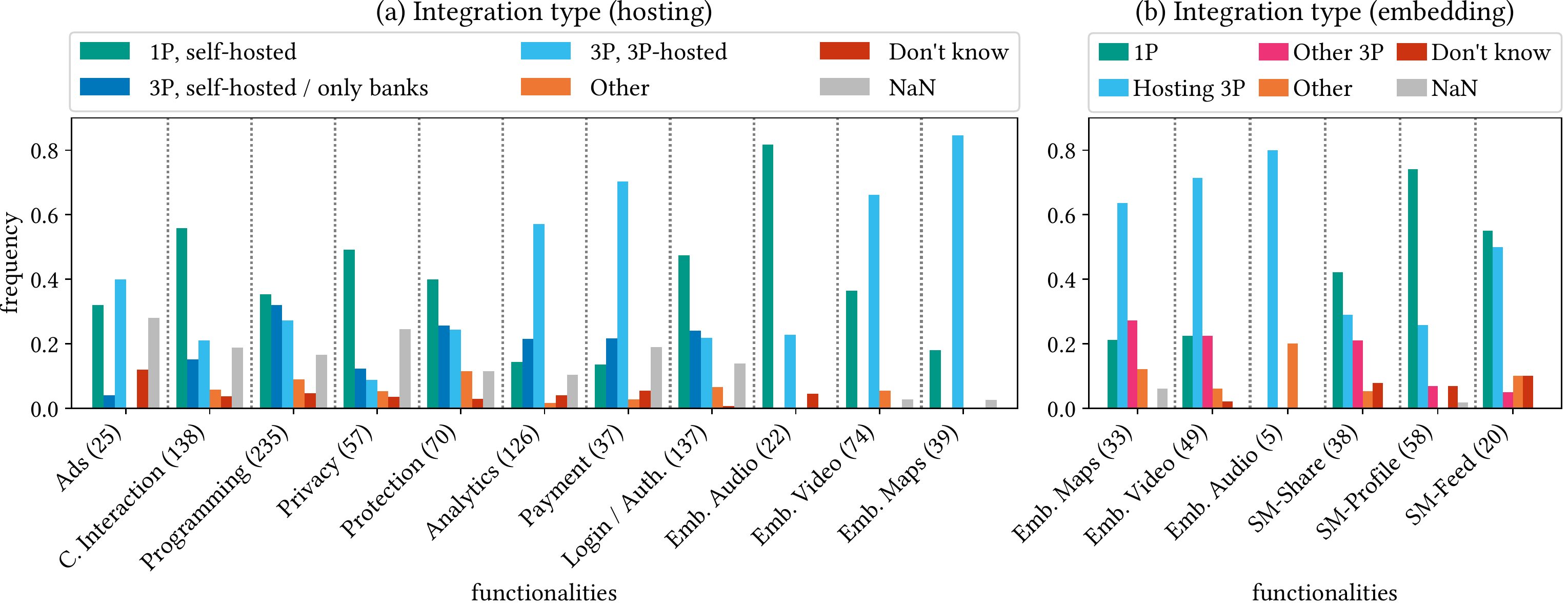}
     \caption{Integration type (Q3-2) for different website functionalities. Left: use of first- vs. third-party hosting; right: source of embedding code for embedded media and social media integration. 
     Numbers are relative to how often the respective
     question had been displayed (see the survey logic in Appendix~\ref{sec:appendix-survey}; n values in x-axis labels).
     }
	\label{fig:integrationtypes}
\end{figure*}

Q3-2 investigated how the different functionalities were integrated into websites. We focused on the hosting location (first-party solution, third-party software installed locally on the own system, or third-party service remotely included from vendor's server). For embedded media and social media, we also investigated (Q3-2c/2d) how remote resources were embedded into the website: via self-written code, code provided by the third party, or an embedding method provided by another third party (such as social media plugins that support multiple social media sites). Figure~\ref{fig:integrationtypes} shows the prevalence of each hosting and embedding type. We observe that websites predominantly self-host solutions for customer interaction (user comments, contact forms, chat, etc.), privacy popups and forms, and embedded audio.
Remotely hosted third-party solutions are dominant for analytics, payment, and hosting of embedded video and map content, while prevalence of the different hosting types was more varied in the other categories.

As shown in Figure~\ref{fig:integrationtypes}(b), remotely hosted media are typically embedded using the code provided by the hosting service. Social media share buttons and embedded feeds, whose functionality implies the requirement to access an API provided by the social network, more or equally often use one of the two third-party embedding variants. By contrast, buttons or links to the website's social media profiles, which do not trigger an action specific to the social network, are more frequently integrated via first-party solutions.

Q3-2 also asked participants to specify which concrete service the website used. Coding revealed the following categories of functionalities to have a clear market leader: advertising (Google Ads / AdSense / DoubleClick for Publishers [63.6\,\% participants who used a third party and provided an answer]), analytics (Google Analytics, 65.7\,\%, followed by Matomo, 10.3\,\%), embedded videos (YouTube, 90\,\%), embedded maps (Google Maps, 62.5\,\%). 
We observed a more varied use of third-party services for programming and design resources (top 3: Bootstrap (18.2\,\%), React (17.5\,\%), jQuery (14.7\,\%)). 
For website protection, participants equally often mentioned web security libraries, which they considered self-hosted third-party services, and Google's reCAPTCHA as the most popular remote third-party service (12.1\,\% for both).

Overall, our findings match expectations:  Third-party use seems more prevalent for website functionalities that (mostly) require third parties to be involved, such as payment services or social media integration, or that were deemed to be complex to self-host or implement, such as analytics or video and map resources~\cite{libert_news_2018}. As for the concrete third-party services used, web tracking research has repeatedly identified Google's services to be the most prevalent third-party services on the Web~\cite{englehardt_1million_2016, karaj_whotracksme_2018, urban_adnetworks_2020}. Still, we measured some efforts at privacy-friendly configuration of Google services.

\subsubsection{Decision Process}

Next, we investigated how people had arrived at these solutions to integrate different website functionalities.

\paragraph{People Involved in the Selection Process}

We learned about who was involved in the selection process in two ways. 
For participants involved in the selection of how to integrate a functionality (Q2-7), we evaluated their roles with regard to the website (Q2-1). 
Across all categories, people involved in selection predominantly had technical roles. 
For given roles we also observed higher involvement in the selection of functionalities that closely relate to that role, such as customer support for customer interaction or sales for advertising. 
Q3-8 asked participants not involved in selection who had made that decision. Here participants most frequently referred to developers, with the notable exception of privacy popups or forms, for which the decision often lay with the legal team, data protection officers, or management. This is also the functionality where participants reported the lowest involvement rates (see Table~\ref{tab:categorystats}). 
Figures~\ref{fig:selectroles} and \ref{fig:selectwho} in Appendix~\ref{sec:people-resources} have details for both questions.

\paragraph{Resources Used for Selection}

Across all categories, participants mainly relied on official websites and documentation to select how to integrate a given functionality (Q3-6); also frequently named were the website's team, online articles, and forums.
The same information sources were reported as most commonly consulted in the selection of ad networks for mobile apps~\cite{mhaidli_ad_networks_2019}. Also confirming the findings of previous work~\cite{mhaidli_ad_networks_2019, balebako_behaviors_2014}, terms of service or privacy policies were rarely consulted, except for payment, privacy plugins, and advertising (16.7\,\% for each). 
Figure~\ref{fig:selectresources} in Appendix~\ref{sec:people-resources} has detailed numbers. 
This suggests that not even functionality where people directly enter sensitive information, such as customer interaction, prompts developers to look up a third-party service's data processing practices. 
This could be due to the complexity and length of these documents, which reinforces the need that third-party services present their key privacy practices in a condensed, easy to understand, and accessible form~\cite{balebako_behaviors_2014}.

\paragraph{Reasons for the Selection of Existing Solutions}

\begin{figure}[tb]
	\centering
    \includegraphics[width=0.5\textwidth]{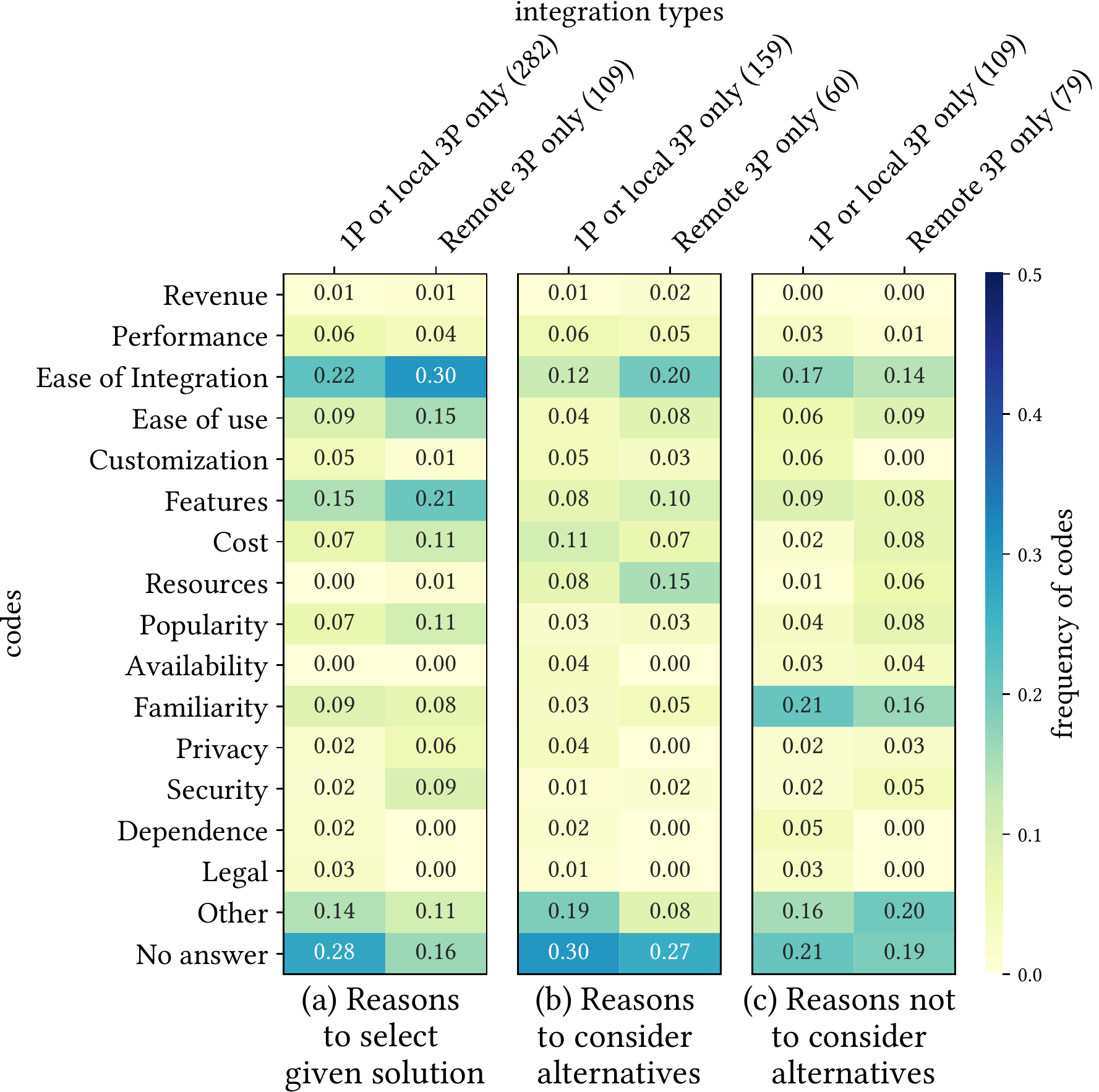}
     \caption{
     Reasons why a given website functionality was integrated in a certain way (a) and why alternatives were considered (b) or not (c), aggregated across functionalities. 
    }
	\label{fig:reasons-heatmaps}
\end{figure}

Coding of the open-ended answers to Q3-3 identified reasons why the respective integration solutions had been selected for each functionality. Figure~\ref{fig:reasons-heatmaps} investigates the reported reasons for two mutually exclusive groups: purely self-hosted solutions, whether first-party or via a locally hosted third party, where collected data is expected to stay on the website's host system, vs. solutions that only rely on remote third-party hosting and thus can involve information being sent to a third-party server. Figure~\ref{fig:reasons-heatmaps}(a) shows the prevalence of each code for each of these integration types, aggregated across all functionalities. We find that the most prevalent decision factors for either integration type are ease of integration and features, though these play a bigger role in the adoption of pure third-party solutions. The ``Other'' category mainly comprises generic answers such as ``I just like it'' (P323-Social) or ``it's the best'' (P188-Login), which explains its relatively high prevalence. 
Beyond these general factors for adoption, we observed that some mainly occurred for certain functionalities, such as revenue for advertising, legal considerations for privacy plugins, security for login/authentication, familiarity for programming/design and analytics, and popularity for payment.
Privacy aspects were rarely mentioned, except for analytics (``I wanted something very minimalistic, non-intrusive'' [P353], ``I care about users privacy'' [P83]). 
These observations confirm findings in the mobile space that third-party adoption is driven by the goal to save time and effort through code reuse~\cite{salza_update_2020}
and additionally finds that these factors can fuel the reasoning both for or against third-party use and there are differences between functionalities.

\paragraph{Consideration of Alternatives}

\begin{figure*}
\centering
\begin{minipage}[t]{.59\textwidth}
  \centering
  \includegraphics[width=1.0\textwidth]{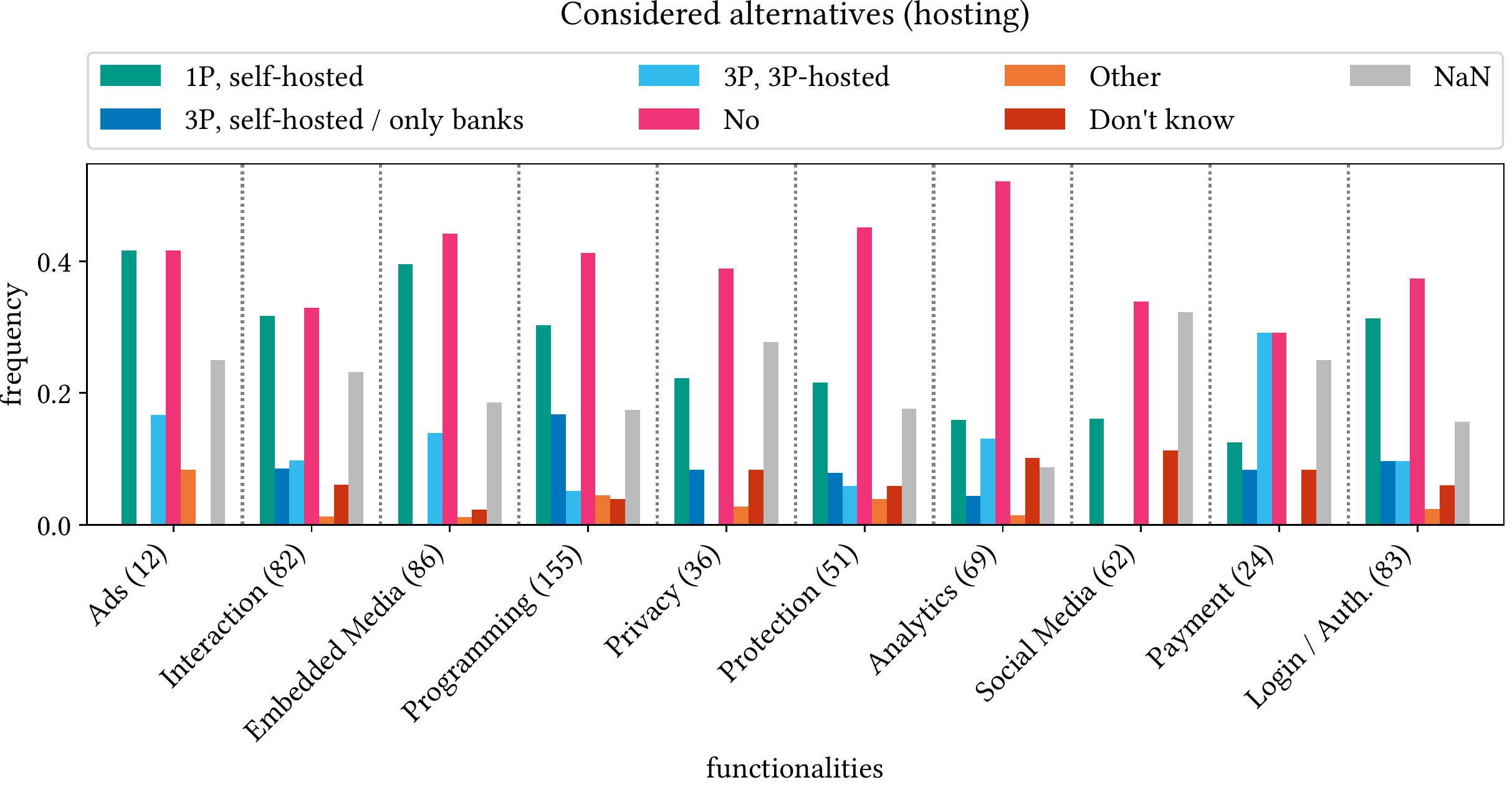}
  \captionof{figure}{Alternatives considered (Q3-4) for the hosting of website functionalities. Numbers are relative to how often the question was displayed (see survey logic in Appendix~\ref{sec:appendix-survey}; n values in x-axis labels).}
  \label{fig:alternatives}
\end{minipage}%
\hfill
\begin{minipage}[t]{.39\textwidth}
  \centering
  \includegraphics[width=1.0\textwidth]{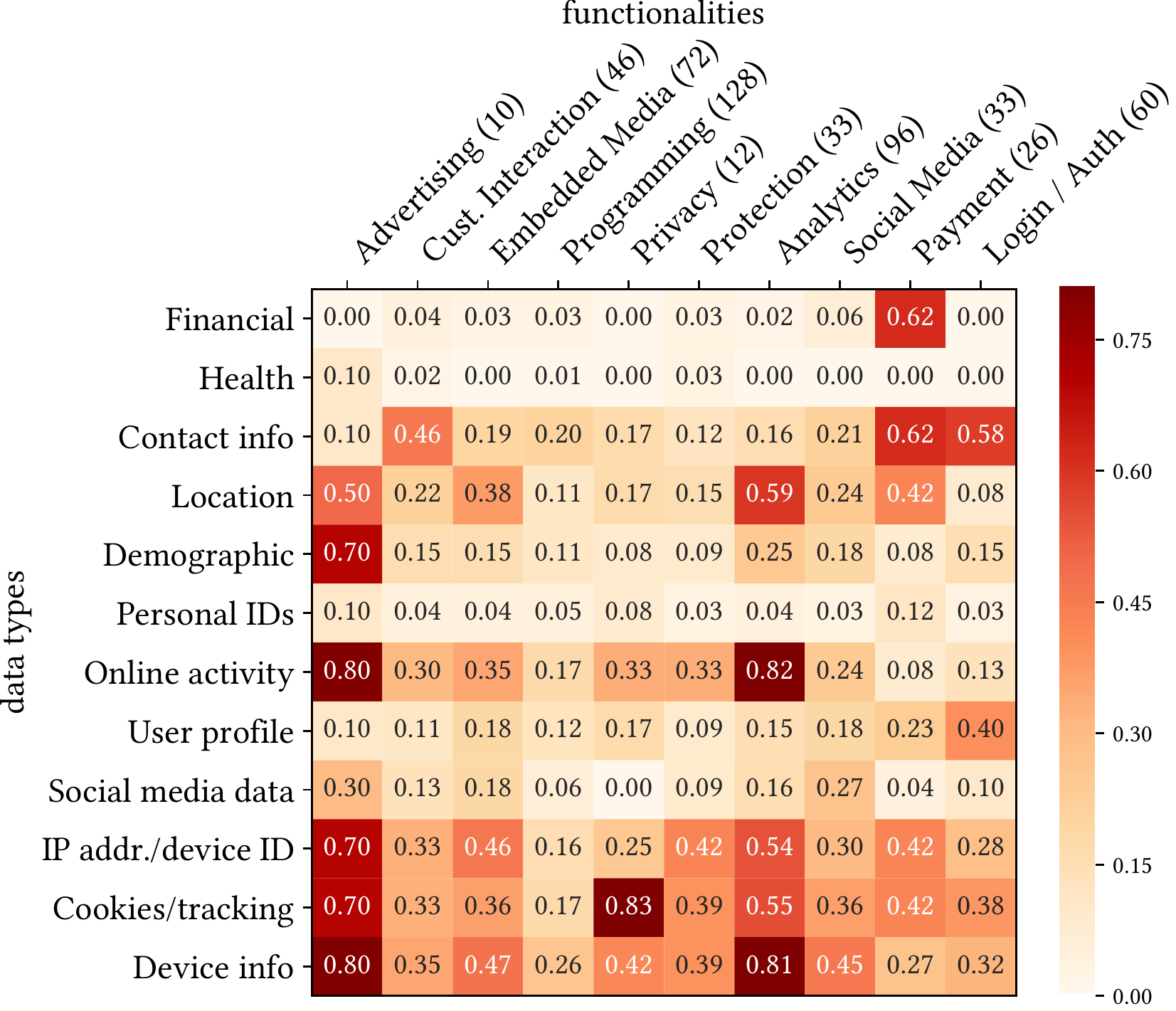}
  \captionof{figure}{Percentage of 3P-using participants who thought what types of personal data the service collected (Q4-1; n values next to func. labels).}
  \label{fig:datacollection-3p}
\end{minipage}
\end{figure*}

Participants involved in the selection of a functionality were asked in Q3-4 whether they had considered alternatives to their chosen integration solution. Figure~\ref{fig:alternatives} shows that across all categories, this was answered negatively by a large share of participants, from 16.7 (advertising) to 50.7\,\% (analytics). 
A similarly low rate was reported in the work of Mhaidli et al., who found only two out of nine interview participants to have made some effort in considering and comparing different mobile ad networks before settling on one~\cite{mhaidli_ad_networks_2019}. Rather, participants were found to select a network based on some ``vague awareness'' of what was popular and commonly used with good experience.
We found similar sentiments in our data for functionality with a clear market leader, notably the prevalent use of analytics, for which the outstanding popularity of Google Analytics was confirmed by our measurements (Table~\ref{tab:privfriendlyhosting}). The answers to Q3-2 suggest that people consider it the ``default'' solution and do not even think about possible alternatives. Except for payment, which is only practical with the involvement of third parties, most considered alternatives were first-party solutions, even for functionalities considered difficult to self-host such as video content or (targeted) advertising~\cite{libert_news_2018}. 
This could again hint at people rarely choosing between different third-party services but rather deciding between either self-implementing a functionality or using a specific third-party service. 

For embedded and social media, participants also had the option to indicate whether they had considered embedding mechanisms from other sources. Of the 62 people who had been asked this question for social media integration, 12 (19.4\,\%) had considered using code provided by the social networks and 4 (6.5\,\%) had considered code by another third party. The embedded media category was shown to 86 participants, 9 of whom (10.5\,\%) had considered self-written embedding code, 3 (3.5\,\%) code provided by the resource-hosting third party, and 4 (4.7\,\%) code by another third party. 

As for the reasons why alternatives were considered or not (Q3-5), Figure~\ref{fig:reasons-heatmaps} in (b) and (c) investigates this for self-hosting vs. pure remote third-party use. We observe that, like for the selection of the current solution (a), ease of integration is a prominent factor to both consider and not to consider alternatives. Somewhat unexpectedly, for pure use of third parties this reason and resources appear to be factors to research rather than to not consider possible alternatives. This could hint towards users of third-party services not always being content with what those offer and decision processes to be complex. However, the most important factor not to consider alternatives appears to be familiarity with the selected solution, for self-hosted solutions even more so than for use of remote third-party services. The ``Other'' responses to this question mainly comprised satisfaction with the current solution, low priority of the respective functionality, or mere statements that it was unnecessary to look for alternatives (``It wasn't required'' [P316-Privacy]; ``The first way worked'' [P241-Analytics]).

\subsection{Privacy Considerations in Integration}

Beyond the selection phase, we investigated participants' privacy practices in the stage of integrating the selected solution.

\subsubsection{Resources Used for Integration}

For integration itself, the answers to Q3-7 paint a similar picture as the resources for selection (Q3-6). Again, the main sources of information were official websites/documentation and the website's team. 
Online articles and forums are less often used for actual integration compared to the selection phase. Terms of service and privacy policies again were rarely consulted. 
Though not directly comparable in answer space, the 20\,\% of privacy plugin users who consulted terms of service or a privacy policy are in the same dimension as the legal information sources used to integrate consent forms for advertising in mobile apps~\cite{tahaei_nudging_2021} (14.1\,\% for ``Legal policies (\eg, GDPR)'' and 9.9\,\% for ``legal teams''). 
Figure~\ref{fig:integresources} in Appendix~\ref{sec:people-resources} shows detailed data for Q3-7.

\subsubsection{Privacy Protection Efforts}

When asked in Q4-2 if they had employed specific measures to protect website visitors' privacy when configuring their solution to implement a functionality, participants' answers did not vary significantly ($ p > 0.05 $, Fisher's exact test) across functionalities. For all of them, about a quarter of participants reported to have employed privacy protection mechanisms, another quarter stated to not have used them, about one third did not know, and the rest did not provide an answer. 

\begin{table*}[tb]
\centering
\caption{
Privacy protection efforts (Q4-2) reported by participants involved in integration or maintenance of a functionality, across all integrations (n = 224). For code definitions see Appendix~\ref{sec:codebooks}.}.
\label{tab:privacyconfigwhat}
\footnotesize
\begin{tabular}{llrr}
\toprule
\textbf{Code} & \textbf{Examples} &  \textbf{n} &   \textbf{\%} \\
\midrule
No personal data  &      ``No user data is logged'' (P337-Analy), ``No personal data is stored'' (P46-Inter) &   9 &   4.0 \\
Data minimization &      ``limited data retention'' (P130-Prote), ``only what we need'' (P1178-Analy) &  38 &  17.0 \\
Self-hosting      &      ``No external service used'' (P190-Priva), ``Coded [it] myself safely'' (P221-Socia) &  19 &   8.5 \\
3P selection      &      ``Remove GA :)'' (P30-Analy), ``non-Google CDNs'' (P855-Progr) &  17 &   7.6 \\
3P setting        &      ``use the no-cookie option'' (P212-Embed), ``anonymize IP on [GA]'' (P1256-Analy) &  26 &  11.6 \\
User consent      &      ``I put them in containers [...] only executed after consent'' (P214-Ads) &  14 &   6.3 \\
Transparency      &      ``privacy policy'' (P535-Ads), ``we follow our privacy policy'' (P66-Inter) &   4 &   1.8 \\
Data access       &      ``access to specific users'' (P955-Progr), ``don't pass any user data'' (P191-Embed) &  18 &   8.0 \\
Anonymization     &      ``anonymus [sic] identifiers'' (P288-Analy), ``obfuscate user ids'' (P917-Inter) &  12 &   5.4 \\
Security          &      ``HTTPS'' (P855-Login), ``password hash'' (P619-Login), ``encryption'' (P1091-Inter) &  34 &  15.2 \\
Other             &      ``too many to list'' (P163-Login), ``look through the [...] source code'' (P695-Embed) &  46 &  20.5 \\
No answer         & Nothing entered, ``1.?????????'' (P352-Login), ``Don't know specifics'' (P53-Socia)      &  29 &  12.9 \\
\bottomrule
\end{tabular}
\end{table*}

Table~\ref{tab:privacyconfigwhat} shows what privacy protection efforts participants reported to have made in the configuration of their solution. Participants frequently referred to data minimization (``I don't really collect user information, and when I do, I keep it to a minimum to get the job done'' [P361-Programming]) and secure transfer (``encryption and [TLS]'' [P84-Inter]).
Another prominent theme in the answers was first- vs. third-party selection, including self-hosting as a means to protect visitors' privacy (``Remove tracking from social media buttons by replacing them with a similar button'' [P385-Social]), careful selection of the third party with privacy in mind (``I chose a font service that I believed would respect user privacy'' [P136-Progr]), and using settings offered by the third-party service to collect less data.
Prominent themes in individual categories are security for login/authentication (32.5\,\%) and customer interaction (28.1\,\%); anonymization, data minimization (22.2\,\% for both), and third-party settings (30.6\,\%) for analytics. 
The explanation for the repeated occurrence of security mechanisms, including access control, is that developers often conflate privacy with security~\cite{tahaei_stackoverflow_2020, hadar_designers_2017}.

Across all categories, only 24 answers to Q4-3a explained the motivation behind the measures to protect visitors' privacy. 20 named regulatory requirements mostly from privacy law but, in the case of payment providers, also industry regulations. Two participants mentioned an unspecified ``requirement'' for analytics and another two a self-commitment to privacy (for analytics and social media). 

\begin{table*}
\centering
\caption{Reasons not to make any specific effort to protect visitors' privacy when integrating or maintaining a functionality, across all integrations (n = 263). 
For code definitions, see Appendix~\ref{sec:codebooks}.}.
\label{tab:privacyconfignot}
\footnotesize
\begin{tabular}{llrr}
\toprule
\textbf{Code} & \textbf{Examples} &  \textbf{n} &   \textbf{\%} \\
\midrule
No data collected &      ``no tracking involved'' (P213-Prote), ``nothing is saved'' (P247-Progr) &  58 &  22.1 \\
Data minimization &      ``We don't ask for anything beyond email address and name'' (P44-Inter) &   8 &   3.0 \\
Self-hosting      &      ``system is on-premise'' (P201-Inter), ``own code without tracking'' (P264-Socia) &  15 &   5.7 \\
Trust in 3P       &      ``The service I used [...] handles security'' (P380-Pay) &  30 &  11.4 \\
Impossible        &      ``no configuration options'' (P11-Ads), ``we are not developing it'' (P32-Prote) &   23 &  8.7 \\
Website purpose   &      ``Internal use only'' (P95-Login), ``page is not ready yet'' (Progr-361) &  26 &   9.9 \\
Priorities        &      ``We [use] analytics to track users. That's the opposite of privacy'' (P290-Analy) &   6 &  2.3 \\
Payoff            &      ``It's more work'' (P132-Analy), ``won't pay back'' (P324-Priva) &   5 &   1.9 \\
Unnecessary       &      ``Why should I'' [P439-Social], ``no need'' (P63-Inter), ``Didn't have to'' (P353-Progr) &  38 &  14.4 \\
Lack of knowledge &      ``I can't understand whole of what [GA] collect[s]'' (P382-Analy) &   2 &  0.8 \\
Other             &      ``Its just a frontend library'' (P338-Progr), ``Existing solutions satisfies'' (P282-Progr) &  16 &   6.1 \\
No answer         &      Nothing entered, ``prefer not to say'' (P91-Ads) &  48 &  18.3 \\
\bottomrule
\end{tabular}
\end{table*}

Table~\ref{tab:privacyconfignot} shows the reported reasons not to make privacy-protecting configurations. Most frequently, the solution was perceived not to collect any personal data, which was especially prevalent for programming/design (39.1\,\%; ``because the third party does not collect anything'' [P109-Progr]), embedded media (18.6\,\%), and social media (34.8\,\%); in the latter case, the responses often referred to first-party integrations of profile buttons or links (``they're just links'' [P98-Socia], ``simply images, wrapped in anchor tags'' [P289-Socia]). Other prominent themes were trust in the third party to adequately protect users' privacy (``I thought the default setup already protects the visitors' privacy enough'' [P243-Analy], ``I trusted [Cloudflare] to not collect excessive information'' [P321-Prote]) and the perception that it was impossible to do anything about collected data (``there is nothing I can do in GA to change the data Google collects'' [P396-Analy]), particularly for analytics (27.6\,\%). 
Trust in third-party vendors and the perceived inability to do something about the data collection were also recurring sentiments in why developers of mobile apps stick to a service's default configuration~\cite{mhaidli_ad_networks_2019}. 
Finally, some answers simply deemed privacy protection unnecessary (``I don't care about privacy because `data is king'{''} [P295-Payment]), prominently for programming/design (39.1\,\%) and embedded media (18.6\,\%).

\subsection{Awareness of Third-Party Data Collection}

Q4-1 more closely investigated the assumed lack of awareness of third-party data collection. 
For the third-party users of each functionality (as by Q3-2), Figure~\ref{fig:datacollection-3p} shows the percentages who thought that the service collected specific types of data.

We observe that participants had a solid understanding of data collection implied by a service's core functionality. For example, a majority of participants reported that third-party privacy popups/forms collect cookies, that payment services require contact and financial information, or that advertising and analytics collect device information and user online activities.
However, beyond this, participants' understanding of data collection was limited. This is especially evident in the case of IP addresses and device information: As HTTP(S) requests to a remote resource involve transmission of a user's IP address and user agent, this information is always available to the third party. More indirect is the opportunity for the third party to derive additional information via these technical parameters, such as tracking users across sites that use that service and learning their browsing behavior. 
It appears that many participants embed third-party software and either do not know or are uncertain of the true extent of data collection by the third party.

This is supported by the responses to Q3-9 that let participants rate the integrated solution with regard to different metrics.
Between 48\,\% (advertising) and 75.71\,\% (website protection) of participants reported to be \textit{Satisfied} or \textit{Very Satisfied} with the privacy offered by their integrated solution, while only up to 8.73\,\% (analytics) expressed some degree of dissatisfaction. This suggests that data collection by third parties is often either accepted or unknown.

\section{Limitations}
\label{sec:limitations}
Our study has some limitations. 
First, we aimed to recruit a diverse sample and we are confident that it provides a wide range of perspectives on third-party adoption but may not include every type of website or third-party user. Websites and third-party services are not easy to categorize, and therefore participants might have interpreted our categories differently (see Table~\ref{tab:3p-categories}). However, we provided examples and aimed for a sensible compromise between lengthy explanations and too much room for interpretation.

Second, a limitation of any survey is self-reported data. 
We cannot verify to what degree participants were actually involved with the provided website or if they consistently answered for the same site.  
Analyzing self-reported information is common in research involving developers~\cite{acar_apis_2017, peixoto_devprivacy_2020, senarath_embedprivacy_2018, mhaidli_ad_networks_2019} and manual inspection of survey responses suggests that participants answered consistently.
Our survey was voluntary and uncompensated, which might have introduced bias, especially since experts tend to be well-paid and hard to reach. However, a lack of compensation was found to yield higher motivation or engagement in developer studies~\cite{acar_apis_2017, acar_github_2017, nadi_apis_2016, gorski_warnings_2018}. 

Further limitations apply to our website analysis. As data collection took multiple weeks, it is possible that in some instances websites changed between participants' responses and website analysis.  
Additional discrepancies might have been introduced due to our categorization differing slightly from WhoTracks.me, third parties using the same domain for multiple purposes, or participants not knowing or naming the functionalities on their website.

\section{Discussion}
\label{sec:discussion}
Our findings provide insights into how web developers and people in similar roles \emph{select} how to integrate a desired functionality, \emph{configure} the selected solution, and if they are \emph{aware} of the privacy risks associated with third-party services.
For selection, we find the prevalence of third-party use to vary by functionality. In configuration, specific efforts to protect website visitors' privacy mostly appear to be made if mandated by technical guidelines on privacy law. 
Based on these findings, we discuss the need to raise awareness of the privacy risks of third-party use on websites and to promote adoption of privacy-friendlier alternatives. 
On the methodological level, our work is a case study for how the perception of research methods previously deemed acceptable can change over time.

\subsection{Lack of Awareness of 3P Data Collection}

Our research confirms the previously suggested lack of awareness~\cite{edri_ethicalwebdev_2020} to what extent the use of third-party functionality on websites can pose risks to visitors' privacy. While developers appear to be aware of data collection closely tied to the main purpose of a third-party service, they often seem to not know or ignore the possibility that their visitors' personal data could be collected for other purposes, or simply trust the third-party service to not collect data or to employ adequate privacy protection.
For analytics, our results hint at a somewhat higher privacy awareness than for other functionalities. This could be due to data collection simply being the main objective of web analytics, or due to prominent and recent guidelines on GDPR-compliant use of web analytics~\cite{strack_anonymizeip_2020, dsk_ga_2020}. 
Similarly, concrete legal requirements have led to the adoption of privacy notices and forms, while developers appear to find it difficult to implement the more generic ``privacy by design'' approach promoted by the the GDPR or the NIST Privacy Framework~\cite{nist_privacyframework_2020}. 
Public discussion and additional guidelines could help raise awareness for the privacy risks of other types of third-party services on websites, and on operationalizing ``privacy by design'' for website development and integration, ideally addressing a wide range of website-related roles. 
Measures to raise awareness would also need to communicate risks beyond the immediate control of developers, as third-party services often connect and share data with each other without users' knowledge~\cite{urban_adnetworks_2020}, and
different understandings of the sensitivity of the collected data, such as IP addresses.

Referring developers to a service's privacy policy is insufficient to communicate its privacy risks. While privacy policies can be expected to contain information about the data collected by a third-party service, our results show that they are rarely used when selecting or configuring services. 
This is unsurprising given that privacy policies are notoriously hard to understand, and the GDPR, a law pursuing greater transparency, has even led to an increase in the length of online privacy policies~\cite{degeling_gdpr_2019}.
As an additional aid, privacy labels for mobile apps have recently been introduced into Apple's and Google's app stores~\cite{apple_privacydetails_2021, frey_googleplaysafety_2021}. With web development not taking place inside such a closed ecosystem, there are no centralized platforms developers could turn to for advice and comparison of different services that integrate a given functionality. For those who use common CMSes, their plugin repositories could introduce similar labels, placing privacy information more prominently than in a legal document. Alternatively, IDEs~\cite{li_coconut_2018} and CMS editors could help assess the number of third-party requests in website code or problematic configurations for popular services and display advice.

\subsection{Promoting Privacy Engineering}

Our work confirms earlier findings from the mobile ad ecosystem that developers often feel resigned and unable to effect change in a third-party ecosystem governed by the exchange of revenue or functionality for access to website visitor's personal information: Previous work found sentiments that users' personal data would be collected by platforms and vendors, irrespective of the developer's decisions~\cite{mhaidli_ad_networks_2019}, and both developers~\cite{mhaidli_ad_networks_2019} and third-party vendors~\cite{tahaei_devsresponsible_2021} deem the respective other party responsible for the protection of users' data. One option to break this cycle of blame and instigate change would be to encourage developers that they can indeed make a difference through privacy-conscious integration of functionality~\cite{mhaidli_ad_networks_2019}; after all, it is developers and end users that made these vendors that prevalent and powerful through use and promotion of their services. 
While in the past it was often browser vendors and developers of privacy-enhancing extensions who fueled advancements in website visitors' privacy, such as the option to block third-party cookies, relegating privacy protection to the browser comes at the risk of breaking websites and could overwhelm users with configuration options and prompts. 
Thus, promoting privacy-by-design with website creators would be a more holistic approach that can ensure that privacy is considered from the beginning of the web development process, desired website functionality works as expected, and the burden is not placed on website visitors. 
A website that practices data minimization and privacy by design could even render annoying consent notices unnecessary for the benefit of both websites and visitors.

We found notable involvement of DPOs or legal experts only for privacy popups or forms, \ie, functionality added for the administration of a website's existing data processing practices. 
This could be an indicator that privacy is still regarded as something to be ``added later'' instead of being considered throughout the development process. Moreover, web development is often done in small teams or by single persons without a privacy professional at hand. 
When the decision is in the hands of developers and made in early stages of the development process, our results show that ease of integration and familiarity with solutions are the driving factors for adoption. This does not necessarily mean that developers do not care about privacy, but it is simply not an important concern given deadlines and limited resources in small teams~\cite{loser_hygiene_2014}. 
While at the beginning of development it is often unclear what user data the final (web) application will need~\cite{balebako_behaviors_2014}, this does not preclude the involvement of privacy considerations from the beginning. Iterative privacy impact and risk assessment processes that continuously evaluate functional requirements against privacy implications could help ensure that the desired functionality is implemented using the least amount of personal data, thus complying with frameworks that follow a data minimization or privacy-by-default approach.

\subsection{Promoting Privacy-Friendlier Alternatives}

While advice to self-host~\cite{roberts_staticassets_2019, pollard_googlefonts_2020} or use privacy-friendly alternatives to popular third-party services~\cite{edri_ethicalwebdev_2020} has increased in recent years, we found that only few participants heeded such advice. Others reported not knowing alternatives to the solution they used or did not have the time or resources to look for them. 
This should be interpreted as a challenge to better promote privacy-friendly alternatives for both the developers of these services and the privacy and security research community at large. 
We found ease of integration, features, and cost to be among the most frequently reported factors that cause developers to adopt a certain solution -- requirements currently easiest to satisfy by a service available free of charge that instead monetizes visitor data. It remains a major challenge to reconcile the demand for usability, features, and lowest possible cost if monetization of visitor data is not an option.

On the configuration level, privacy-friendlier options do exist but are often hidden or obscured by dark patterns~\cite{tahaei_appdevs_2022}. For example, YouTube's setting for ``privacy-enhanced mode'' is only revealed when one scrolls down in the ``Embed'' dialog while the standard embed code is directly visible. 
Vendors could encourage use of the privacy-friendly configuration by making it more prominent or even the default, though there is no incentive for this if the service's business model is based on monetization of personal data, as is often the case with third-party services offered free of monetary cost. Privacy laws and court rulings were identified as drivers of privacy-related settings in ad networks ~\cite{tahaei_devsresponsible_2021}, analytics services~\cite{noyb_austriaga_2022, pollet_cnilga_2022}, and cookie consent notices~\cite{ecj_planet49_2019}.
Thus, public policy measures and regulatory guidance could go one step further and require vendors to make the privacy-friendly option the default.

\subsection{Methodological Implications}
\label{sec:discussion-ethics}

Section~\ref{sec:ethics} described how recruitment via email addresses in public GitHub commit metadata came under the scrutiny of our state's data protection authority. We now discuss what part of the process had raised concerns with the DPA, what this means for future recruitment in privacy and security research, and what could be done in advance to decrease the likelihood of facing similar problems.

\subsubsection{Recruiting Developers on GitHub}

Email recipients who asked how we found their email address on GitHub often pointed out that they had set their email address to ``private'' on their GitHub user profiles. While this setting hides the address from the public profile, it does not affect the visibility of the email address in commits to public GitHub repositories. 
Any given commit into a public repository has a corresponding \texttt{*.patch} file, available at \url{https://github.com/<user>/<repository>/commit/<commit\_hash>.patch}. The second line in this file shows the author of the commit, along with their email address. This is due to the core concept behind GitHub's public repositories, where all commits, including metadata, are public. 
The documentation~\cite{github_commitemail_2022} describes how users can configure Git(Hub) to use their GitHub-provided ``noreply'' email address, which will remove their real email address from the commit metadata but still associate their contributions with their GitHub account.

Email feedback showed that many GitHub users are not aware of these mechanics and settings. This was also the issue at the core of the DPA's assessment, which argued that GitHub users pushing commits into public repositories did not expect to be contacted via their commit email addresses for the purpose of scientific research, and this lack of awareness constituted a legitimate interest of the user that outweighed public interest in scientific research. 
In addition, users of GitHub's API are bound by GitHub's terms of service and privacy statement~\cite{github_termsofservice_2022}. 
GitHub's privacy policy considers a user email address public information (unless made private as described above) but proceeds to limit its use ``for the purpose for which [the] user authorized it''~\cite{github_privacypolicy_2022}. Following the DPA's argument, this likely does not include being contacted for the purpose of participation in scientific research. 
It remains for the community to decide what influence such company policies should have on the question of what is considered ethical in privacy and security research, and, looking further ahead, how to handle company policies on data use that contradict what is permissible under applicable law.

For future recruitment of study participants we recommend, as also suggested by the DPA, to only use contact information that has visibly been made public by the individuals themselves with the intention of allowing the general public to contact them.
GitHub's email address mechanics and users' lack of knowledge about them had neither been mentioned nor addressed by previous work that used public GitHub repositories for recruitment. 
We hope that our experience can inform the ongoing debate about ethics in privacy and security research and the search for alternatives to reach diverse sets of developers in a reliable, ethical, and affordable way.

\subsubsection{The Need for A Priori Community-Based Ethics Review}

It has long been best practice in human subjects research to obtain prior review via an institutional review board (IRB) or a similar entity to ensure that participation in the study does not cause undue harm to humans. However, in practice, many institutions, especially outside the US, do not have such a review board, or review is not always mandatory, as was the case for our study. 
But even if prior IRB review had been available, it remains doubtful whether it could have prevented the complaint to the DPA. The main goal of IRB review is to ensure that a study complies with human subjects regulations, not to provide a comprehensive ethics and legal assessment. In fact, we took additional steps to get GDPR assessments from our institutions' DPOs before running the study. 
The challenge is that in privacy and security research, a deep ethics and legal review would often require specific technical domain knowledge
(\eg, GitHub's handling of commit email addresses), associated risks, and their legal evaluation. These are aspects that are often not covered by IRB guidelines or board members' background due to their differing function. 
Legal assessment in particular can be subject to rapid evolution through new laws and court rulings, requiring involvement of legal experts who keep up with this constant change. 

Recently the privacy and security research community has identified this need for thorough ethical review and multiple venues have set up ethics committees that can be involved in the review process if a submission raises ethical concerns with reviewers. This work went through this very process, and we highly value the thorough ethics review we received, which concluded that we adequately addressed our study's ethical implications. While ethical review after submission is an important step in ensuring that published privacy and security research did not cause undue harm to the people whose behavior and systems were studied, it effectively comes too late, at a time any potential harm would have already been caused. Hence, the community needs to consider how to provide ethical guidance before potentially harmful research is carried out, for example, by means of a ``standing ethics review board'' of expert volunteers that can complement institutional review in the study design phase. Such a priori ethics review would (1) help prevent unethical privacy and security research before it occurs, (2) provide researchers with experience and confidence in how to address ethical implications, and (3) minimize the sometimes arbitrary and ad-hoc assessments of a study's ethical implications by reviewers. An existing example is the Tor Research Safety Board~\cite{torproject_safetyboard_2022}; providing  committees of domain experts that cover the whole privacy and security field would pose a major challenge. 
Hence, such a priori review would not have to be mandatory for all submissions but could become a valued community resource.

\section{Conclusion}
\label{sec:conclusion}
We report findings from an online survey with 395 
people working with websites on how common website functionalities are implemented, in particular if third-party services are used and whether and how respective privacy implications have been considered.

While we observe that the selection process is influenced by a variety of factors, we find that often factors such as a third-party service's popularity and ease of integration fuel adoption decisions. By contrast, website visitors' privacy only plays a notable role in web analytics, a functional category which has been explicitly addressed by data protection authorities. Except for privacy popups and forms, data protection officers and legal counsels are rarely involved in the decision processes that lead to the integration of third-party services into websites despite potential privacy implications. 

\newpage

\begin{acks}
We would like to thank the anonymous reviewers for the detailed feedback and the comprehensive ethics assessment, and our shepherd for their guidance to further improve this work. This research was funded by the MKW-NRW Research Training Groups SecHuman and NERD.NRW.
\end{acks}

\bibliographystyle{ACM-Reference-Format}
\bibliography{bibliography}
\balance

\appendix
\clearpage
\section{Survey}
\label{sec:appendix-survey}
\small

This appendix presents the main part of the survey, \ie, without the intro text, privacy policy, debriefing, and end message. Except for Q2-0 in the GitHub--Mandatory condition, all questions were non-mandatory.

\subsection*{Survey Title}

Web Technologies: Selection, Integration, and Configuration

\subsection*{1. Your Background}
First we would like to learn about your background and your work on websites. Throughout this survey, by ``work on websites'' we mean your involvement to some degree in the design, development, deployment, maintenance, and/or management of a website.

\begin{compactenum}
    \item[1-1] How many websites have you worked on in the last 3 years?  
    [single choice, answer options: 0, 1, 2--5, 6--10, 11--25, 26--50, 51--100, $ > $ 100]
\end{compactenum}

\begin{compactenum}
    \item[1-2] What is your current employment status with regard to your work on websites? [multiple choice] 
    \begin{compactitem}
    \setlength{\itemindent}{-1em}
        \item Full-time employment
        \item Part-time employment
        \item Self-employed / freelancer
        \item Intern
        \item Hobbyist
        \item Unemployed
        \item Retired
        \item Unable to work
        \item Other: [free text]
        \item Prefer not to say
    \end{compactitem}
\end{compactenum}

\begin{compactenum}
    \item[1-3] Below is a list of functionalities often found on websites. Which of these functionalities have you previously worked with on websites? [multiple choice; order of answers randomized]  
    \begin{compactitem}
    \setlength{\itemindent}{-1em}
        \item Advertising (\eg, banner ads, video ads, content recommendation, affiliate links)
        \item Customer / user interaction (\eg, user comments, contact forms, chat, mailing lists)
        \item Embedded media (\eg, video, audio, maps, slideshows)
        \item Front-end libraries or design resources (\eg, non-standard fonts, CSS frameworks, JavaScript libraries)
        \item User login / authentication
        \item Payment systems
        \item Privacy popups / privacy forms (\eg, cookie consent notices, CCPA ``Do not sell'')
        \item Website protection (\eg, anti-spam, bot mitigation techniques)
        \item Social media integration (e.g., social media buttons, widgets, embedded feeds) 
        \item Web analytics (\eg, page visits, heatmaps, session replay)
    \end{compactitem}
\end{compactenum}

\subsubsection*{2a. Website}

To learn more about your experience with different web technologies, the rest of the survey will ask you about a specific website you have recently worked on.

\begin{compactenum}
    \item[2-0] 
Please name one website you recently worked on, \ie, you were involved in the design, development, deployment, maintenance, or management of that website, and that you remember well.

(\textbf{If recruited via website:} Ideally, this is the website through which we contacted you, which is mentioned in the email invitation to this survey. If you were not in any way involved in the design, development, deployment, maintenance, or management of that website, you are welcome to provide another website you recently worked on.)

We will keep this website -- and any other information that could identify you -- confidential and only share it with involved researchers.

Please enter the website's web address below, including the top-level domain (\eg, \texttt{youtube.com}, \texttt{guardian.co.uk}).

For the remainder of this survey, all questions are going to refer to this website as ``the website.'' [free text]

\textit{(In the GitHub--Mandatory condition, we required participants to enter something but did not check if it was a valid URL.)}

\end{compactenum}

\subsubsection*{2b. Website Info}

In Part 2 of the survey, we would like to learn some more information about the website you just named.

\begin{compactenum}
    \item[2-1] What is / are your role(s) with regard to the website? [multiple choice] 
    \begin{compactitem}
    \setlength{\itemindent}{-1em}
        \item Product or project manager
        \item Content creator or contributor
        \item Social media manager
        \item Marketing
        \item Sales
        \item Quality assurance 
        \item User experience
        \item (Web) developer, programmer, or software engineer
        \item Administrator or (web) operator
        \item Legal counsel
        \item Data protection officer
        \item Customer service / customer support / customer relations
        \item Other: [free text]
   \end{compactitem}
\end{compactenum}

\begin{compactenum}
    \item[2-2] What is roughly the size of the team working on the website, \ie, how many people have been involved in the website's design, development, deployment, maintenance, and management? 
    [single choice, answer options: I am the only team member, 2--5, 6--10, 11--25, 26--50, 51--100, $ > $ 100, Don't know]
\end{compactenum}
    
\begin{compactenum}
    \item[2-3] Please select which country the company or organization operating the website is based in. If the company or organization has sites in multiple countries, please select the country in which the company or organization's headquarters are located. [single choice, answer options: dropdown list with names of all countries]
\end{compactenum}

\begin{compactenum}
    \item[2-4] What regions or countries is the website targeting or being used in? 
    [free text]
\end{compactenum}

\begin{compactenum}
    \item[2-5] What is the website's revenue model? [multiple choice] 
    \begin{compactitem}
    \setlength{\itemindent}{-1em}
        \item Targeted advertising (\eg, ad networks)
        \item Non-targeted advertising (\eg, contextual or static ads)
        \item Affiliate marketing / affiliate links 
        \item Donations 
        \item Subscriptions / membership 
        \item Sponsored posts / articles 
        \item Products / services sold on the website 
        \item Supported by other revenue streams (i.e., goods or services not directly sold on the website)
        \item Other: [free text]
        \item Not applicable (website does not have a revenue model)
        \item Don't know 
    \end{compactitem}
\end{compactenum}

\begin{compactenum}
    \item[2-6] Which of the following features or functionalities are used on the website? [single choice for each, answer options: Yes / No / Not sure] 
    \begin{compactitem}
    \setlength{\itemindent}{-1em}
        \item Advertising (\eg, banner ads, video ads, content recommendation, affiliate links)
        \item Customer / user interaction (\eg, user comments, contact forms, chat, mailing lists)
        \item Embedded media (\eg, video, audio, maps, slideshows)
        \item Front-end libraries or design resources (\eg, non-standard fonts, CSS frameworks, JavaScript libraries)
        \item User login / authentication
        \item Payment systems
        \item Privacy popups / privacy forms (\eg, cookie consent notices, CCPA ``Do not sell'')
        \item Website protection (\eg, anti-spam, bot mitigation techniques)
        \item Social media integration (\eg, social media buttons, widgets, embedded feeds)
        \item Web analytics (\eg, page visits, heatmaps, session replay)
    \end{compactitem}
\end{compactenum}
    
\begin{compactenum}
    \item[2-7] 
    For each of the following functionalities present on the website, how involved have you been regarding their integration into the website? 
    [list of all functionalities tagged with ``Yes'' in previous question, single choice for each, answer options:]
    \begin{compactitem}
    \setlength{\itemindent}{-1em}
        \item I decided how to integrate this functionality
        \item I integrated / implemented this functionality
        \item I maintain or manage the integration of this functionality
        \item I have not been involved in the integration of this functionality
    \end{compactitem}
\end{compactenum}

\subsection*{3. Integration of Website Functionalities (category-specific)}

In Part 3 we would like to ask you a few questions about the integration of some of the functionalities you indicated to have worked with on the website. You will be shown these questions for at most three different functionalities, regardless of how many you have selected in the previous question.

\noindent\emph{(For up to three categories randomly selected from those the participant has indicated involvement in the previous question, they are asked the following questions.)}

\noindent You indicated that you have been involved to some degree in the integration of [FUNCTIONALITY (examples)] on the website. Now we would like to ask you a few more questions about how this functionality has been integrated. 

\begin{compactenum}
    \item[3-1] For which purposes or use cases is [FUNCTIONALITY] technology used on the website?
    [free text] 
\end{compactenum}

\begin{compactenum}
\item[3-2a.] 

\begin{compactitem}
\setlength{\itemindent}{-1em}
\item[] \textbf{(Generic:)}
Which technology has been used to integrate [FUNCTIONALITY] into the website? If the website uses multiple technologies for this, please consider all of them combined (your ``solution'') when answering the following questions. [multiple choice + free text]
\begin{compactitem}
    \setlength{\itemindent}{-1em}
    \item We developed it ourselves  
    \item We installed a third-party software on the website's host system (please name software:)
    \item We integrated an external third-party service (please name service:)
    \item Other (please specify):
    \item Don't know
\end{compactitem}

\item[b.]  \textbf{(Payment:)} What kind of payment service(s) does the website use? [multiple choice + free text]
\begin{compactitem}
    \setlength{\itemindent}{-1em}
    \item Payment method(s) that do not require other parties for processing (\eg, cash, gift cards) (please name method(s):)
    \item Service(s) that only involve banks on either side (\eg, bank transfer, Lastschrift) (please name service(s):)
    \item Service(s) that involve third parties (\eg, credit card, PayPal) (please name service(s):)
    \item Other (please specify:)
    \item Don't know
\end{compactitem}

\item[c.] 
\textbf{(Embedded Media:)} 
\begin{compactitem}
\setlength{\itemindent}{-1em}
\item[i.]

What type of embedded media does the website use? [multiple choice]
\begin{compactitem}
\setlength{\itemindent}{-1em}
    \item Embedded maps
    \item Embedded videos
    \item Embedded audio
    \item Other (please specify:) [free text]
    \item Don't know
\end{compactitem}

\item[ii.] 
\begin{compactitem}
    \item[(1)] \textbf{(If map, audio, or video:)} You indicated that the website uses embedded (maps | videos | audio).

    \begin{compactenum}
    \setlength{\itemindent}{-1em}
        \item[(a)] Where are these (map | video | audio) resources hosted? [multiple choice]
        \begin{compactitem}
        \setlength{\itemindent}{-1em}
            \item The (map | video | audio) resources are hosted on the website's host system
            \item The (map | video | audio) resources are hosted with a third-party service (please name service:) [free text]
            \item Other (please specify:) [free text]
            \item Don't know
        \end{compactitem}

        \item[(b)] 
        \textbf{(If map, audio, or video and third-party hosting:)} How are these externally hosted (map | video | audio resources) embedded into the website? If the website uses multiple technologies for this, please consider all of them combined (your ``solution'') when answering the following questions. [multiple choice]
        \begin{compactitem}
        \setlength{\itemindent}{-1em}
            \item Embedding code provided by the third party that hosts the resources
            \item Embedding code provided by another third-party service (please specify service:) [free text]
            \item Embedding code we have written ourselves
            \item Other (please specify:) [free text]
            \item Don't know
        \end{compactitem}
    \end{compactenum}

\item[(2)] 
\textbf{(If ``Other'':)} You indicated that the website uses some other kind of embedded content. How is this content integrated into the website? If the website uses multiple technologies for this, please consider all of them combined (your ``solution'') when answering the following questions. [free text]
\end{compactitem}
\end{compactitem}

\item[d.] 
\textbf{(Social Media:)}
\begin{compactitem}
\setlength{\itemindent}{-1em}
\item[i.] 
 What type of social media integration does the website use? [multiple choice]
\begin{compactitem}
\setlength{\itemindent}{-1em}
    \item Profile buttons or links
    \item Share  buttons or widgets
    \item Embedded posts or feeds
    \item Other: [free text]
    \item Don't know
\end{compactitem}

\item[ii.] 
\begin{compactitem}
    \item[(1)] \textbf{(If profile / share buttons or embedded:)} 
    You indicated that the website uses (buttons or links to social media profiles | social media share buttons or widgets | embedded social media posts or feeds). How are they integrated into the website? Which technology has been used to integrate them into the website? If the website uses multiple technologies for this, please consider all of them combined (your ``solution'') when answering the following questions. [multiple choice]
\begin{compactitem}
\setlength{\itemindent}{-1em}
    \item Code we have written ourselves
    \item Code provided by social media site(s) 
    \item Code or plugin provided by another third-party service (please specify service:) [free text]
    \item Other (please specify:) [free text]
    \item Don't know
\end{compactitem}
    \item[(2)] \textbf{(If ``Other'':)} 
    You indicated that the website uses some other kind of social media integration. How is it integrated into the website? If the website uses multiple technologies for this, please consider all of them combined (your ``solution'') when answering the following questions. [free text]
    \end{compactitem}
\end{compactitem}
\end{compactitem}
\end{compactenum}

\begin{compactenum}
    \item[3-3] \textbf{(If involved in selection:)} You indicated that you were involved in deciding how [FUNCTIONALITY] was integrated into the website. 
    Please describe why this specific type of integration or this particular service was selected. [free text] 
\end{compactenum}    

\begin{compactenum}
    \item[3-4] \textbf{(If involved in selection:)} 
    \begin{compactitem}
    \setlength{\itemindent}{-1em}
        \item[a.]  \textbf{(Generic:)} When making this decision, were other ways for integrating [FUNCTIONALITY] into the website considered? [multiple choice]
    
        \begin{compactitem}
        \setlength{\itemindent}{-1em}
            \item We considered a solution we have developed (or were going to develop) ourselves
            \item We considered (another) third-party software installed on the website's host system (please name software:) [free text]
            \item We considered a(nother) service hosted with a third party (please name service(s):) [free text]
            \item We directly decided to use the current solution
            \item Other (please specify:) [free text]
            \item Don't know
        \end{compactitem}
    \end{compactitem}
    
    \begin{compactitem}
    \setlength{\itemindent}{-1em}
    \item[b.]  \textbf{(Payment:)} 
    When making this decision, were other ways for integrating payment systems into the website considered? [multiple choice]
    \begin{compactitem}
    \setlength{\itemindent}{-1em}
        \item We considered (other) methods that do not include any other party (\eg, cash, gift cards) (please name method(s):) [free text]
        \item We considered service(s) that only involve banks on either side (please name service(s):) [free text]
        \item We considered (other) service(s) that involve third parties (please name service(s):) [free text]
        \item We directly decided to use the current solution
        \item Other (please specify:) [free text]
        \item Don't know
    \end{compactitem}
    \end{compactitem}
    
    \begin{compactitem}
    \setlength{\itemindent}{-1em}
    \item[c.]  \textbf{(Embedded Media:)} When making this decision were other ways for integrating embedded media into the website considered? [multiple choice]
    \begin{compactitem}
    \setlength{\itemindent}{-1em}
        \item We considered self-hosting the embedded media resources
        \item We considered hosting the embedded media resources with a(nother) third party (please specify service:) [free text]
        \item We considered embedding code provided by the third-party service that hosts the resources (please specify service:) [free text]
        \item We considered embedding code provided by a different third-party service (please specify service:) [free text]
        \item We considered embedding code we have written (or were going to write) ourselves
        \item We directly decided to use the current solution
        \item Other (please specify:) [free text]
        \item Don't know
    \end{compactitem}
    \end{compactitem}
    
    \begin{compactitem}
    \setlength{\itemindent}{-1em}
    \item[d.]  \textbf{(Social Media:)} When making this decision, were other ways for integrating social media into the website considered? [multiple choice]
    \begin{compactitem}
    \setlength{\itemindent}{-1em}
        \item We considered a solution we have developed (or were going to develop) ourselves
        \item We considered code provided by the social media site(s)
        \item We considered a solution provided by a different third-party service (please specify service:) [free text]
        \item We directly decided to use the current solution
        \item Other (please specify:) [free text]
        \item Don't know
    \end{compactitem}
    \end{compactitem}
\end{compactenum}

\begin{compactenum}
    \item[3-5]  \textbf{(If involved in selection:)} Why were other ways to integrate [FUNCTIONALITY] into the website (not) considered?
    [free text] 
\end{compactenum}

\begin{compactenum}
    \item[3-6]  \textbf{(If involved in selection:)} 
    Which sources of information did you use to select a solution to integrate [FUNCTIONALITY] into the website? [multiple choice]
    \begin{compactitem}
    \setlength{\itemindent}{-1em}
        \item The website's team 
        \item Professional network (people external to the website team)
        \item Private network (\eg, friends)
        \item Sales representative of third-party software / service
        \item Official website(s) / documentation of third-party software / service
        \item Legal documents by third-party software / service (\eg, terms of service, privacy policy)
        \item Online blogs / magazine articles
        \item Online discussion forums (\eg, Reddit, StackOverflow)
        \item Other: [free text]
    \end{compactitem}
\end{compactenum}

\begin{compactenum}
    \item[3-7] \textbf{(If involved in implementation or maintenance:)} 
    Which sources of information did you use to configure the [FUNCTIONALITY] solution on the website? [multiple choice, same answer options as in Q3-6]
\end{compactenum} 

\begin{compactenum} 
    \item[3-8] \textbf{(If not involved in selection:)} You indicated that you were not involved in the decision how to integrate [FUNCTIONALITY] into the website. Who decided how [FUNCTIONALITY] should be integrated into the website? [multiple choice]
    \begin{compactitem}
    \setlength{\itemindent}{-1em}
        \item Product or project manager(s)
        \item Content creator(s) or contributor(s)
        \item Social media manager(s)
        \item Marketing
        \item Sales
        \item Quality assurance
        \item User experience
        \item (Web) developer(s), programmer(s), or software engineer(s)
        \item Administrator(s) or (web) operator(s)
        \item Legal counsel(s)
        \item Data protection officer(s)
        \item Customer service / customer support / customer relations
		\item CEO and/or other upper level management
		\item Investor(s)
        \item Other: [free text]
        \item Don't know
    \end{compactitem}
\end{compactenum}

\begin{compactenum}
    \item[3-9] Overall, how satisfied are you with the [FUNCTIONALITY] integration solution on the website, with regard to the following criteria? [single choice for each of the following, answer options: Very satisfied, Satisfied, Neither satisfied nor dissatisfied, Dissatisfied, Very dissatisfied, Don't know] 
    \begin{compactitem}
    \setlength{\itemindent}{-1em}
        \item Visitors' privacy
        \item Ease of integration
        \item Ease of use for visitors
        \item Performance (\eg, page speed)
        \item Features meet requirements
    \end{compactitem}
\end{compactenum}

\subsection*{4. Data Practices of Website Functionalities (category-specific)}

In Part 4 of the survey, we would like to learn more about your experience with the data practices of the technologies we just asked you about in Part 3.

\noindent\textit{(The following questions are asked for each functionality for which the participant has also seen Part 3.)}

\begin{compactenum}
    \item[4-1] 
    \textbf{(If third-party service is used to implement [FUNCTIONALITY]:)}
    Sometimes third-party services, when integrated into a website, collect information about the website's visitors, either to provide the service or for their own/other purposes. 
    To the best of your knowledge, what information about the website's visitors does the third-party solution used for [FUNCTIONALITY] collect?
    
    [Items taken from the ``Information Type'' section of the annotation scheme for the OPP-115 corpus of privacy policies~\cite{wilson_opp115_2016}; single choice for each, answer options: Yes, No, Unsure]
   \begin{compactitem}
        \setlength{\itemindent}{-1em}
        \item Financial information (\eg, credit or debit card data, credit scores)
        \item Health, genetic, or biometric data
        \item Contact information (\eg, name, email address, phone number)
        \item Location (\eg, GPS location, postal code)
        \item Demographic data (\eg, gender, age, education)
        \item Personal identifiers (\eg, social security, ID card or driver's license number)
        \item User online activities (\eg, pages visited, time spent on pages)
        \item User profile on the website (\eg, profile settings, data the user has uploaded to the website)
        \item Social media data
        \item IP address or device IDs
        \item Cookies or other tracking elements
        \item Device information (\eg, browser or operating system used by website visitors)
    \end{compactitem}
\end{compactenum}

\begin{compactenum}
    \item[4-2]  \textbf{(If involved in implementation or maintenance:)} 
    Did you make any specific effort(s) to protect the website's visitors' privacy when configuring the [FUNCTIONALITY] solution on the website? [single choice]
    \begin{compactitem}
        \setlength{\itemindent}{-1em}
        \item Yes
        \item No
        \item Don't know
    \end{compactitem}
    
    \item[4-3a.] 
    \begin{compactenum}
    \setlength{\itemindent}{-1em}
    \item[]  \textbf{(If yes:)} Please describe which efforts you have made and why. [free text]
    
    \item[b.]  \textbf{(If no:)} Please describe why you did not make any specific efforts. [free text]
    \end{compactenum}
    
\end{compactenum}

\subsection*{5. Demographics}

Finally, we would like to ask you some basic demographic questions to better understand who participated in our study.

\begin{compactenum}
    \item[5-1] What is your age (in years)? [single choice]
    [18--24, 25--34, 35--44, 45--54, 55--64, 65--74, 75+, Prefer not to disclose]
\end{compactenum}

\begin{compactenum}
    \item[5-2] What is your gender?\footnote{As recommended by Spiel et al.~\cite{spiel_gender_2019}.} [multiple choice]
    \begin{compactitem}
    \setlength{\itemindent}{-1em}
        \item Woman
        \item Man
        \item Nonbinary
        \item Prefer to self-describe: [free text]
        \item Prefer not to disclose
    \end{compactitem}
\end{compactenum}

\begin{compactenum}
    \item[5-3] What is the highest educational degree you have completed? [single choice]
    \begin{compactitem}
    \setlength{\itemindent}{-1em}
        \item No schooling completed
        \item Some high school, no diploma
        \item High school graduate, diploma, or equivalent (\eg, GED, Abitur, baccalaur{\'e}at)
        \item Some college credit, no degree
        \item Trade / technical / vocational training
        \item Associate degree
        \item Bachelor's degree
        \item Master's degree or equivalent (\eg, German Diplom)
        \item Professional degree (\eg, JD, MD, German Staatsexamen)
        \item Doctoral degree (\eg, PhD)
        \item Other: [free text]
        \item Prefer not to disclose
    \end{compactitem}
\end{compactenum}

\begin{compactenum}
    \item[5-4] In what field(s) did you receive your degree or vocational training?\footnote{Adapted from a Pew Research survey~\cite{pew_highered_2013}, using the subcategories for some fields.} [multiple choice]
    \begin{compactitem}
    \setlength{\itemindent}{-1em}
        \item Computer and information sciences
        \item Mathematics
        \item Engineering
        \item Life sciences (\eg, biology, health sciences, medicine)
        \item Social sciences / social work / human services
        \item Education
        \item Law
        \item Psychology / behavioral science
        \item Business / economics
        \item Liberal arts / humanities
        \item Art / music
        \item Journalism
        \item Vocational
        \item Other: [free text]
        \item Not applicable
        \item Prefer not to disclose
    \end{compactitem}
\end{compactenum}

\begin{compactenum}
    \item[5-5] 
    Have you ever received any kind of training or educated yourself on data protection or privacy? [single choice]
    \begin{compactitem}
    \setlength{\itemindent}{-1em}
        \item Yes (please specify:) [free text]
        \item No
        \item Prefer not to disclose
    \end{compactitem}
\end{compactenum}

\clearpage
\section{Codebooks}
\label{sec:codebooks}

\subsection{Reasons For/Against Certain Solutions to Integrate a Functionality (Q3-3/Q3-5)}

\begin{compactdesc}
    \item[Revenue]              (Not) using this solution affects revenue and conversion, and therefore income.
    \item[Performance]          (Not) using this service affects site performance, \eg, loading times or server computation load.
    \item[Ease of Integration]  It is very easy/hard to implement or integrate the solution.
    \item[Ease of Use]          It is very easy/hard to use the solution (once it has been integrated).
    \item[Customization]        The solution can(not) be easily customized to the participant's needs.
    \item[Features]             The solution (does not) offer(s) specific features that the participant deems important for their use case.
    \item[Cost]                 It would be cheap/expensive to use the solution.
    \item[Resources]            The solution was cheap in non-monetary resources, such as time or workforce.
    \item[Popularity]           The solution is very popular, widespread, or even a market leader.
    \item[Availability]         The solution is easily accessible, \eg, because it is already in use.
    \item[Familiarity]          Friends, colleagues, or the participant themselves know or use the service, allowing the participant to benefit from this experience.
    \item[Privacy]              Privacy was a relevant reason; the service was used because it, \eg, allowed privacy-increasing configurations.
    \item[Security]             Security was a relevant reason; the service was used because it, \eg, allowed security-increasing configurations.
    \item[Dependence]           (In)dependence on/from libraries or services that, \eg, might suffer from outages or be abandoned by their developers in the future.
    \item[Legal]                The service was used due to legal requirements to, \eg, add a privacy policy or cookie banner.
    \item[Other]                Other concrete reasons not covered by the codes above.
    \item[No answer]            The participant did not provide an answer to the question, either by filling in nothing, something incomprehensible, or not providing an answer to the question (\eg, instead repeating what they did, not why).
\end{compactdesc}

\subsection{Type of Effort Made to Protect Website Visitors' Privacy (Q4-3a)}

\begin{compactdesc}
    \item[No Personal Data]     No personal data is collected.
    \item[Data Minimization]    Only the necessary personal data is collected; data collection is as minimal as possible.
    \item[Self-Hosting]         Services are self-hosted; all data stays within the respective organization.
    \item[3P Selection]         Third-party services are carefully selected; there was a conscious decision for/against certain third parties. 
    \item[3P Setting]           Third-party services are configured in ways that increase privacy, \eg, by limiting the amount of collected data, encrypting data etc.
    \item[User Consent]         Users were informed that their data would be available to third parties and gave their consented to this data processing before the functionality was loaded. 
    \item[Transparency]         Privacy Policies or similar information on data practices is available to users.
    \item[Data Access]          The access to the data/server is limited; access is controlled.
    \item[Anonymization]        Data is anonymized and cannot be used to identify certain individuals.
    \item[Security]             Security practices to avoid known attacks or vulnerabilities (\eg, to avoid XSS) are in place, that increase privacy by decreasing the probability of data leaks.
    \item[Other]                Other concrete reasons not covered by the codes above.
    \item[No answer]            The participant did not provide an answer to the question, either by filling in nothing, something incomprehensible, or not providing an answer to the question.
\end{compactdesc}

\subsection{Reasons to Protect Website Visitors' Privacy (Q4-3a)}

\begin{compactdesc}
    \item[Regulatory] Some regulatory framework, \eg, law or industry standards, mandate privacy protection measures.
    \item[Requirement] An unspecified requirement, \eg, by the customer, mandates privacy protection measures.
    \item[Self-Commitment] The participant applied privacy protection measures out of intrinsic motivation, without external influence.
\end{compactdesc}

\subsection{Reasons Not to Protect Website Visitors' Privacy (Q4-3b)}

\begin{compactdesc}
    \item[No Data Collected]    The solution does not collect any personal data, so there is no need for privacy protection.
    \item[Data Minimization]    Only strictly necessary data is collected, so there was/is no need for privacy protection.
    \item[Self-Hosting]         The service is self-hosted, and there is no need for additional measures as access is limited and no external services are involved.
    \item[Trust in 3P]          Trust in the third party to employ adequate measures to protect visitors' privacy. 
    \item[Impossible]           Data collection cannot be controlled or limited, it is impossible to increase privacy.
    \item[Website Purpose]      The website's purpose makes privacy protection unnecessary, \eg, because its main content is only accessible in a logged-in state. 
    \item[Priorities]           Functionality (by adding third party services) has a higher priority than increasing privacy by avoiding these services.
    \item[Payoff]               Privacy measures include too much effort in terms of \eg, workload, cost, time.
    \item[Unnecessary]          It is not necessary to increase privacy. Answers with this code include no explanation, but often indicate a lack of awareness, care or external requirements.
    \item[Lack of Knowledge]    Participants are not able to adjust settings due to \eg, a lack of knowledge or skill with the service.
    \item[Other]                Other concrete reasons not covered by the codes above.
    \item[No answer]            The participant did not provide an answer to the question, either by filling in nothing, something incomprehensible, or not providing an answer to the question.
\end{compactdesc}

\balance

\onecolumn{
\section{People and Resources Involved in Selection and Integration}
\label{sec:people-resources}

\begin{figure*}[htbp]
\centering
\begin{minipage}[t]{.49\textwidth}
  \centering
  \includegraphics[width=1.0\textwidth]{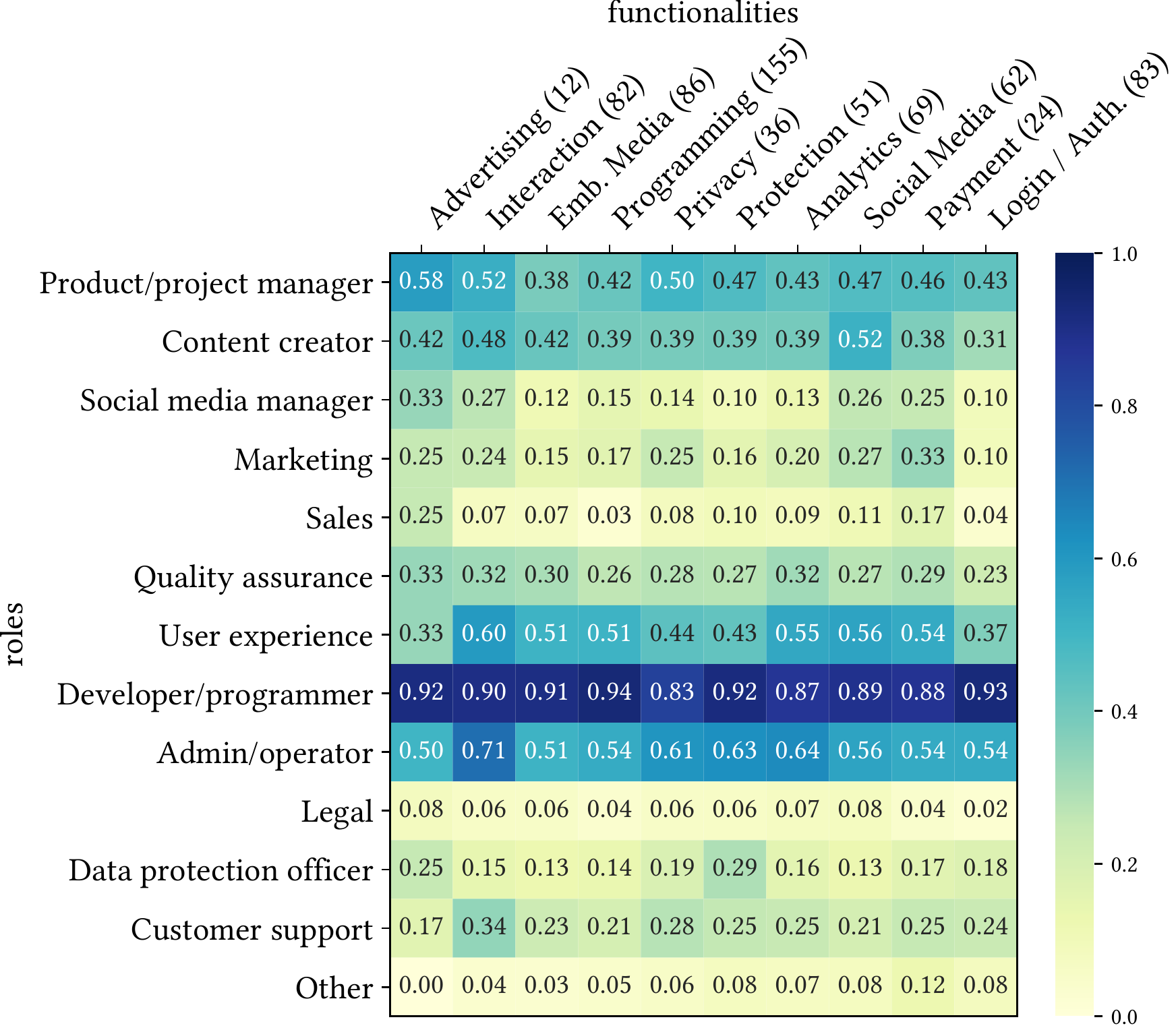}
  \captionof{figure}{Responsibility for selection: for participants involved in the selection of a given functionality (Q2-7), their roles in relation to the website (Q2-1).
     }
  \label{fig:selectroles}
\end{minipage}%
\hfill
\begin{minipage}[t]{.49\textwidth}
  \centering
  \includegraphics[width=1.0\textwidth]{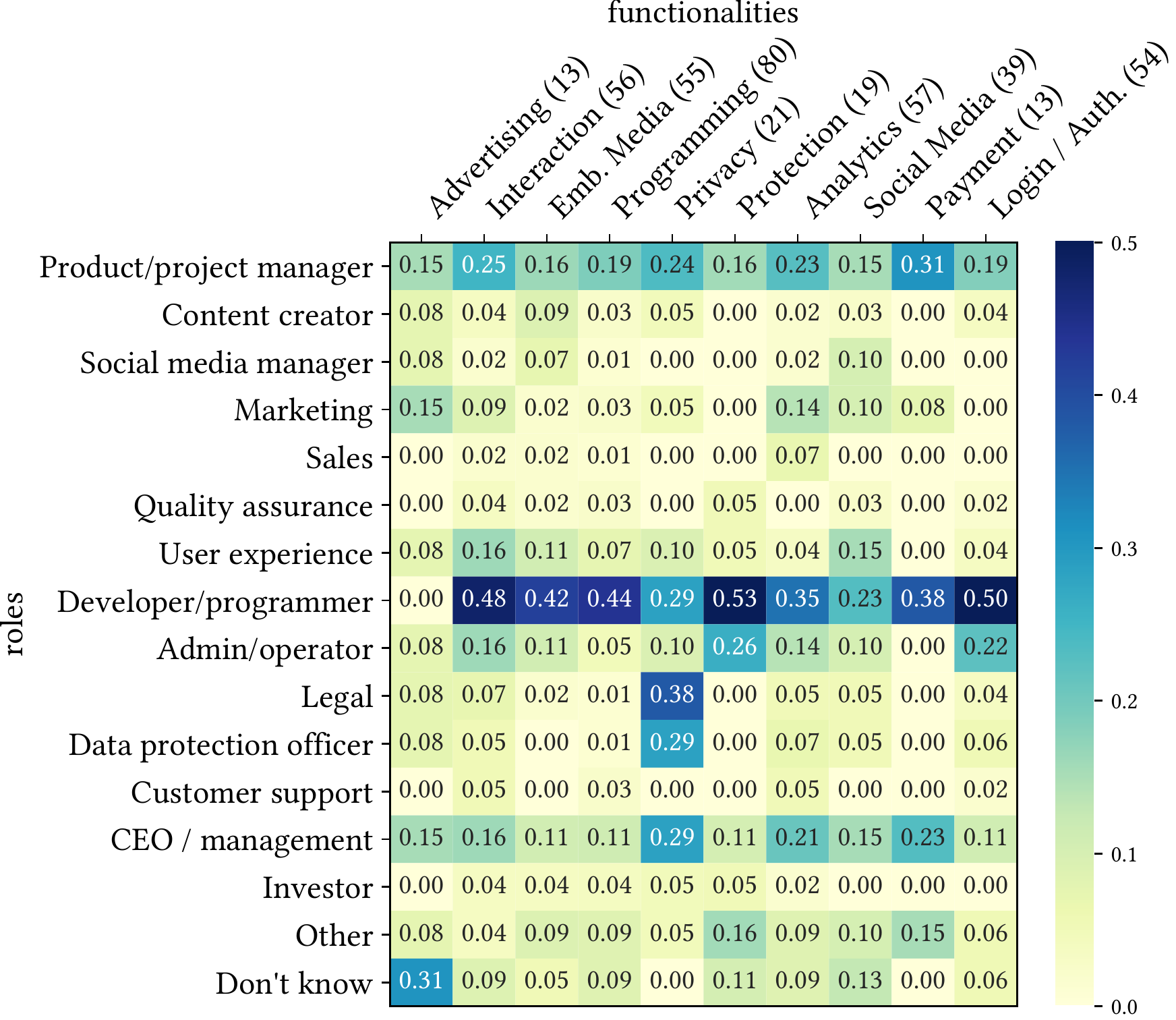}
  \captionof{figure}{Responsibility for selection: for participants not involved in selection, who was responsible (Q3-8).
  }
  \label{fig:selectwho}
\end{minipage}
\end{figure*}
}

\begin{figure*}[htbp]
    \centering
     \includegraphics[width=1.0\textwidth]{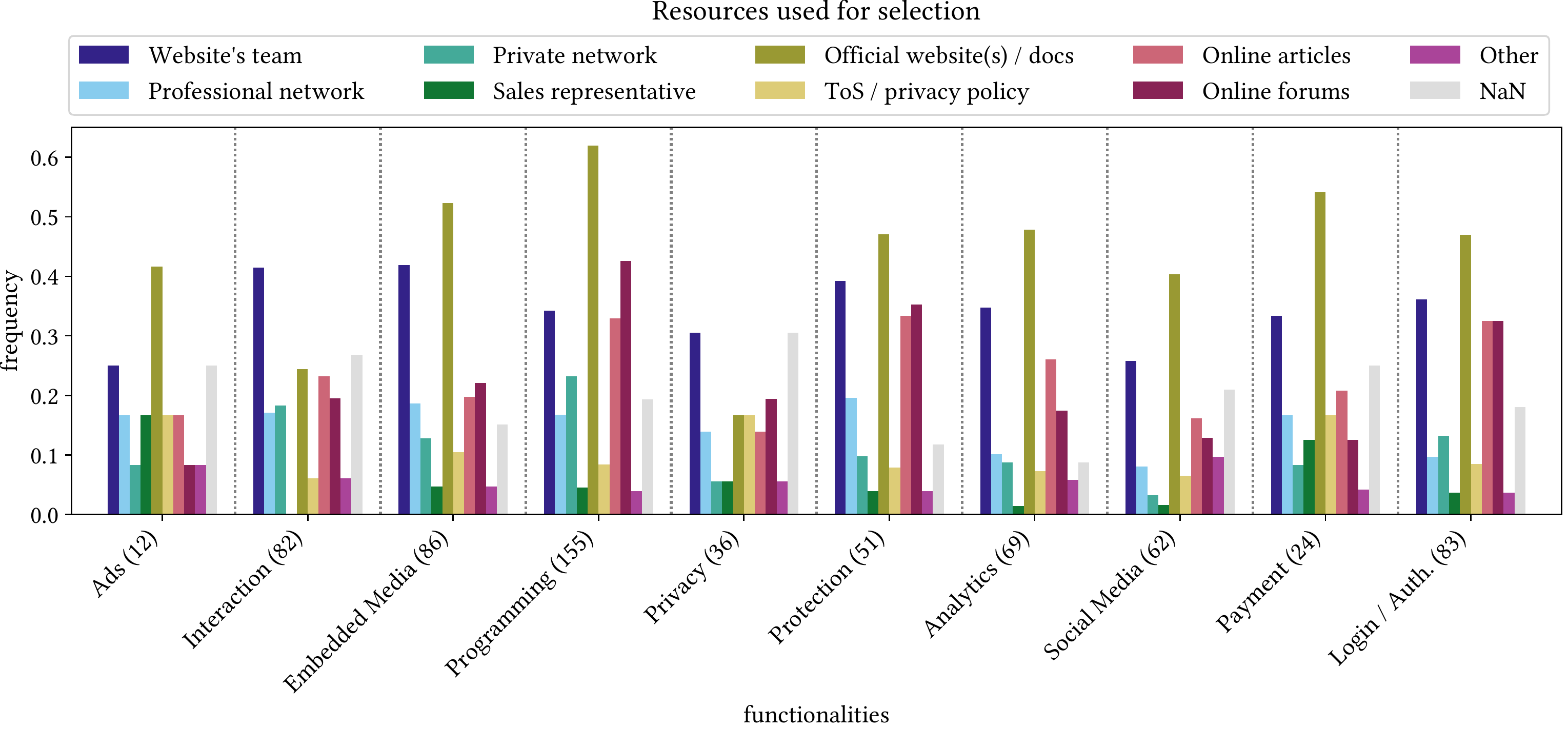}
     \caption{Resources used to select how to integrate a website functionality (Q3-6). Numbers are relative to the people involved in selection of the respective functionality, shown in the x-axis labels.}
	\label{fig:selectresources}
\end{figure*}

\begin{figure*}[t]
    \centering
    \includegraphics[width=1.0\textwidth]{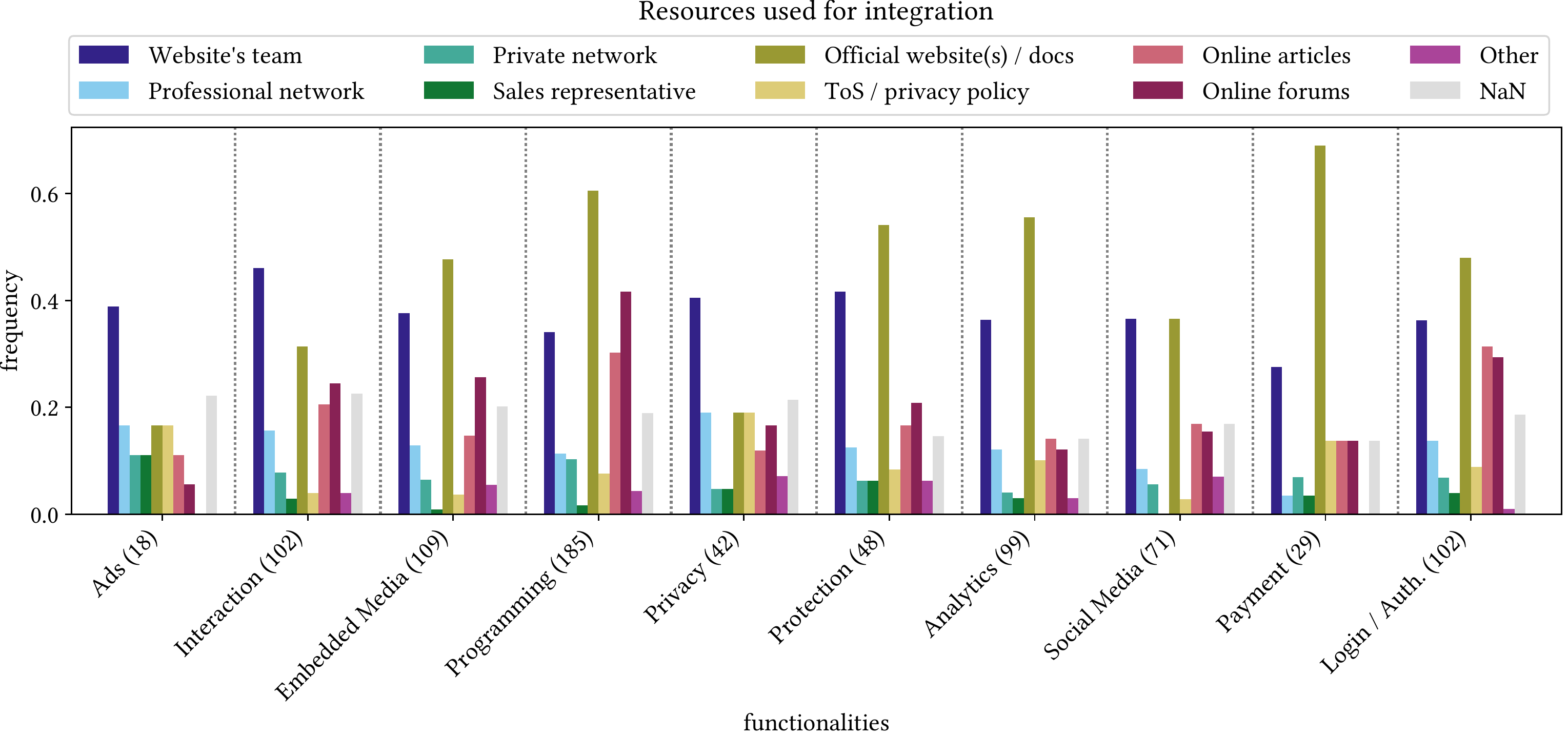}
     \captionof{figure}{Resources used in the integration of a website functionality (Q3-7). Numbers are relative to the people involved in integration or maintenance of the respective functionality, shown in the x-axis labels.}
	\label{fig:integresources}
\end{figure*}

\clearpage
\section{Participant \& Website Statistics}
\label{sec:sample-stats}



\begin{table*}[hbt!]
\small
\caption[Participant Demographics]{\label{tab:participant-stats}
Participants' demographics (Part 5 of the survey) and background (Part 1 of the survey, Q2-1, and Q2-2). $^{\text{C}} $ indicates coded open-ended answers, $^{\text{M}} $ indicates multiple-choice questions or multiply assigned codes for which (response) counts can sum up to more than 100\,\%. Percentage values are relative to the total number of survey responses ($ n = 395 $). For the coded open-ended answers to the type of privacy training received (Q5-5; bottom left, indented list), percentage values are relative to to the number of participants who indicated to have received prior privacy training ($ n = 166 $).
}
\begin{minipage}{0.48\textwidth}
\renewcommand{\thefootnote}{\thempfootnote}
\renewcommand{\arraystretch}{0.75}
\begin{tabular*}{\textwidth}{@{}>{\bfseries}ll@{\extracolsep{\fill}}*{2}{r}}
\toprule
\multicolumn{4}{c}{\textbf{Demographics}} \\
& & \multicolumn{1}{c}{\textbf{n}} & \multicolumn{1}{c}{\textbf{\%}} \\

\midrule
\multirow{8}{*}{\rotatebox[origin=c]{90}{Age}}
& 18--24                            &   132    &  33.4  \\
& 25--34                            &   121    &  30.6  \\
& 35--44                            &    76    &  19.2  \\
& 45--54                            &    30    &   7.6  \\
& 55--64                            &    20    &   5.1  \\
& 65--74                            &     5    &   1.3  \\
& 75+                               &     1    &   0.3  \\
& N/A                               &    10    &   2.6  \\

\midrule\multirow{5}{*}{\rotatebox[origin=c]{90}{Gender$^{\text{M}} $}}
& Woman                            &    40    &  10.1  \\
& Man                              &   336    &  85.1  \\
& Nonbinary                        &     4    &   1.0  \\
& Self-described                   &     3    &   0.8  \\
& N/A                              &    13    &   3.3  \\
\midrule\multirow{12}{*}{\rotatebox[origin=c]{90}{Education}}
& No schooling completed                    &     5  &   1.3  \\
& Some high school, no diploma              &    14  &   3.5  \\
& High school graduate                      &    57  &  14.4  \\
& Some college credit, no degree            &    39  &   9.9  \\
& Trade / technical / vocat. training   &   13  &   3.3  \\
& Associate degree                          &    5  &  1.3  \\
& Bachelor's degree                         &   139 &  35.2  \\
& Master's degree                           &    77 &  19.5  \\
& Professional degree                       &     9 &   2.3  \\
& Doctoral degree                           &    21 &   5.3  \\
& Other                                     &     4 &   1.0  \\
& N/A                                       &     2 &   0.5  \\
\midrule\multirow{17}{*}{\rotatebox[origin=c]{90}{Field of Degree$^{\text{M}} $}}
& Computer \& information sciences  &   222 &  56.2 \\
& Mathematics                       &    53 &  13.4 \\
& Engineering                       &    89 &  22.5 \\
& Life sciences                     &    19 &   4.8 \\
& Physical sciences                 &    26 &   6.6 \\
& Social sciences                   &    23 &   5.8 \\
& Education                         &    19 &   4.8 \\
& Law                               &     2 &   0.5 \\
& Psychology                        &     5 &   1.3 \\
& Business / economics              &    41 &  10.4 \\
& Liberal arts / humanities         &    23 &   5.8 \\
& Art / music                       &    10 &   2.5 \\
& Journalism                        &     7 &   1.8 \\
& Vocational                        &     3 &   0.8 \\
& Not applicable                    &    24 &   6.1 \\
& Other                             &     9 &   2.3 \\
& N/A                               &    12 &   3.0 \\
\midrule\multirow{12}{*}{\rotatebox[origin=c]{90}{Privacy Training$^{\text{C, M}} $}}
& Yes                               &   166 &   42.0 \\
\cmidrule{2-4}
& \hspace{3mm}Self-taught                       &    64 &   38.6 \\
& \hspace{3mm}Employer training                 &    39 &   23.5 \\
& \hspace{3mm}`Learning by doing'               &    10 &    6.0 \\
& \hspace{3mm}University / school               &    18 &   10.8 \\
& \hspace{3mm}Online courses                    &    11 &   6.6 \\
& \hspace{3mm}Other courses                     &    25 &   15.1 \\
& \hspace{3mm}Professional network              &     7 &    4.2 \\
& \hspace{3mm}Other                             &     5 &    3.0 \\
& \hspace{3mm}N/A                               &    15 &    9.0 \\
\cmidrule{2-4}
& No                                &   189 &   47.8 \\
& N/A                               &    40 &   10.1 \\
\bottomrule
\end{tabular*}
\end{minipage}
\hfill
\begin{minipage}{0.48\textwidth}
\vspace{-3.95cm}
\renewcommand{\thefootnote}{\thempfootnote}
\renewcommand{\arraystretch}{0.75}
\begin{tabular*}{\textwidth}{@{}>{\bfseries}ll@{\extracolsep{\fill}}*{2}{r}}
\toprule
\multicolumn{4}{c}{\textbf{Background}} \\
& & \multicolumn{1}{c}{\textbf{n}} & \multicolumn{1}{c}{\textbf{\%}} \\

\midrule
\multirow{8}{*}{\rotatebox[origin=c]{90}{\# Websites}} 
& 1                                 &    18    &   4.6  \\
& 2--5                              &   173    &  43.8  \\
& 6--10                             &   107    &  27.1  \\
& 11--25                            &    47    &  11.9  \\
& 26--50                            &    29    &   7.3  \\
& 51--100                           &    10    &   2.5  \\
& > 100                             &    10    &   2.5  \\
& N/A                               &     1    &   0.3  \\

\midrule\multirow{9}{*}{\rotatebox[origin=c]{90}{Employm. Type$ ^{\text{M}} $}} 
& Full-time employment                      &   165  &  41.8  \\
& Part-time employment                      &    49  &  12.4  \\
& Self-employment / freelancer              &   130  &  32.9  \\
& Intern                                    &    30  &   7.6  \\
& Student                                   &    15  &   3.8  \\
& Hobbyist                                  &   124  &  31.4  \\
& Unemployed                                &    39  &   9.9  \\
& Retired                                   &     3  &   0.8  \\
& Other                                     &     6 &    1.5  \\

\midrule\multirow{11}{*}{\rotatebox[origin=c]{90}{Exp. w. Functionality$ ^{\text{M}} $}} 
& Advertising                   &    91 &  23.0 \\
& Analytics                     &   215 &  54.4 \\
& Customer interaction          &   293 &  74.2 \\
& Embedded media                &   258 &  65.3 \\
& User login / authentication   &   318 &  80.5 \\
& Payment                       &   129 &  32.7 \\
& Programming / design          &   328 &  83.0 \\
& Privacy popups / forms        &   118 &  29.9 \\
& Social media integration      &   204 &  51.6 \\
& Website protection            &   130 &  32.9 \\
& N/A                           &     1 &   0.3 \\

\midrule\multirow{13}{*}{\rotatebox[origin=c]{90}{Role(s) with Website$ ^{\text{M}} $}} 
& Product / project manager         &   136 &  34.4 \\
& Content creator / contributor     &   142 &  35.9 \\
& Social media manager              &    51 &  12.9 \\
& Marketing                         &    63 &  15.9 \\
& Sales                             &    19 &   4.8 \\
& Quality assurance                 &    93 &  23.5 \\
& User experience                   &   162 &  41.0 \\
& (Web) developer etc.              &   337 &  85.3 \\
& Administrator / (web) operator    &   194 &  49.1 \\
& Legal counsel                     &    13 &   3.3 \\
& Data protection officer           &    43 &  10.9 \\
& Customer support / relations      &    71 &  18.0 \\
& Other                             &    19 &   4.8 \\

\bottomrule
\end{tabular*}
\end{minipage}
\end{table*}



\begin{table*}[htb!]
\small
\centering
\caption[Website Statistics]{\label{tab:website-stats}
Statistics about the self-selected websites participants considered while answering the survey. $^{\text{C}} $ indicates coded open-ended answers, $^{\text{M}} $ indicates multiple-choice questions or multiply assigned codes or tags for which (response) counts can sum up to more than 100\,\%. Statistics in the left column are from Part 2 of the survey and percentage values are relative to the total number of survey responses ($ n = 395 $). Statistics in the right column result from the analysis of the website URLs provided by participants in Q2-0 and percentage values are relative to the number of unique entered domains ($ n = 361 $). For the most frequently occurring TLDs, subdomains on popular hosting platforms such as github.io or herokuapp.com, used by 57 sites, were considered distinct TLDs. 
. 
}
\begin{minipage}{0.48\textwidth}
\renewcommand{\thefootnote}{\thempfootnote}
\renewcommand{\arraystretch}{0.75}
\begin{tabular*}{\textwidth}{@{}>{\bfseries}ll@{\extracolsep{\fill}}*{6}{r}}
\toprule
\multicolumn{4}{c}{\textbf{Survey Responses}} \\
& & \multicolumn{1}{c}{\textbf{n}} & \multicolumn{1}{c}{\textbf{\%}} \\

\midrule\multirow{8}{*}{\rotatebox[origin=c]{90}{Team Size}} 
& I am the only team member         &   145    &  36.7  \\
& 2--5                              &   141    &  35.7  \\
& 6--10                             &    50    &  12.7  \\
& 11--25                            &    36    &   9.1  \\
& 26--50                            &     5    &   1.3  \\
& 51--100                           &     5    &   1.3  \\
& > 100                             &    10    &   2.5  \\
& Don't know                        &     3    &   0.8  \\

\midrule\multirow{12}{*}{\rotatebox[origin=c]{90}{Country of Website HQ}} 
& United States of America                  &    70  &  17.7  \\
& Germany                                   &    46  &  11.6  \\
& United Kingdom                            &    21  &   5.3  \\
& Russia                                    &    20  &   5.1  \\
& Brazil                                    &    18  &   4.6  \\
& India                                     &    15  &   3.8  \\
& China                                     &    13  &   3.3  \\
& Switzerland                               &    12  &   3.0  \\
& Canada                                    &    11 &    2.8  \\
& The Netherlands                           &    11 &    2.8  \\
& Other                                     &   154 &   39.0  \\
& N/A                                       &     4 &    1.0  \\

\midrule\multirow{12}{*}{\rotatebox[origin=c]{90}{Target Region/Audience$^{\text{C}} $}} 
& Global                                    &   128  &  32.4  \\
& Europe                                    &    56  &  14.2  \\
& Multiple regions                          &    30  &   7.6  \\
& United States of America                  &    26  &   6.6  \\
& East Asia                                 &    17  &   4.3  \\
& Brazil                                    &    15  &   3.8  \\
& Southeast Asia                            &    15  &   3.8  \\
& Africa                                    &    12  &   3.0  \\
& Russia / CIS                              &    12  &   3.0  \\
& North America                     &    11  &   2.8  \\
& Other                                     &    20  &   5.1  \\
& N/A                                       &    53  &  13.4  \\

\midrule\multirow{12}{*}{\rotatebox[origin=c]{90}{Revenue model$^{\text{M}} $}} 
& Targeted advertising                &     32 &      8.1 \\
& Non-targeted advertising            &     22 &      5.6 \\
& Affiliate marketing / links         &     21 &      5.3 \\
& Donations                           &     37 &      9.4 \\
& Subscriptions / membership          &     69 &     17.5 \\
& Sponsored posts / articles          &     22 &      5.6 \\
& Products / services sold on website &     81 &     20.5 \\
& Other revenue streams               &     57 &     14.4 \\
& Not applicable / no revenue model   &    177 &     44.8 \\
& Don't know                          &      5 &      1.3 \\
& Other                               &     17 &      4.3 \\
& N/A                                 &      2 &      0.5 \\
\bottomrule
\end{tabular*}
\end{minipage}
\hfill
\begin{minipage}{0.48\textwidth}
\vspace{-5.95cm}
\renewcommand{\thefootnote}{\thempfootnote}
\renewcommand{\arraystretch}{0.75}
\begin{tabular*}{\textwidth}{@{}>{\bfseries}ll@{\extracolsep{\fill}}*{6}{r}}
\toprule
\multicolumn{4}{c}{\textbf{Site Analysis}} \\
& & \multicolumn{1}{c}{\textbf{n}} & \multicolumn{1}{c}{\textbf{\%}} \\

\midrule\multirow{12}{*}{\rotatebox[origin=c]{90}{Top-Level Domains}} 
& .com                  &   107  &  29.6  \\
& .org                  &    30  &   8.3  \\
& .de                   &    24  &   6.6  \\
& .github.io            &    19  &   5.3  \\
& .herokuapp.com        &    17  &   4.7  \\
& .dev                  &    12  &   3.3  \\
& .net                  &    11  &   3.0  \\
& .com.br               &    10  &   2.7  \\
& .ru                   &    10  &   2.7  \\
& .io                   &     7  &   1.9  \\

& Other                 &   115  &  31.9  \\

\midrule\multirow{12}{*}{\rotatebox[origin=c]{90}{Website Categories$^{\text{M}} $}} 
& Business                      &    65  &  18.0  \\
& Internet Services             &    54  &  15.0  \\
& Education / Reference         &    38  &  10.5  \\
& Personal Pages                &    21  &   5.8  \\
& Software / Hardware           &    19  &   5.3  \\
& Interactive Web Applications  &    18  &   5.0  \\
& Blogs / Wiki                  &    15  &   4.2  \\
& Marketing / Merchandising     &    11  &   3.0  \\
& Finance / Banking             &    10  &   2.8  \\
& Online Shopping               &    10  &   2.8  \\
& Other                 &    129  &   35.7  \\
& Uncategorized                 &    48  &  13.3  \\
\bottomrule

\end{tabular*}
\end{minipage}

\end{table*}

\end{document}